# Hierarchical Bayesian Operational Modal Analysis: Theory and Computations

Omid Sedehi[1], Lambros S. Katafygiotis[2] and Costas Papadimitriou[3*]

**Abstract**

This paper presents a hierarchical Bayesian modeling framework for the uncertainty quantification in modal identification of linear dynamical systems using multiple vibration data sets. This novel framework integrates the state-of-the-art Bayesian formulations into a hierarchical setting aiming to capture both the identification precision and the variability prompted due to modeling errors. Such developments have been absent from the modal identification literature, sustained as a long-standing problem at the research spotlight. Central to this framework is a Gaussian hyper probability model, whose mean and covariance matrix are unknown encapsulating the uncertainty of the modal parameters. Detailed computation of this hierarchical model is addressed under two major algorithms using Markov chain Monte Carlo (MCMC) sampling and Laplace asymptotic approximation methods. Since for a small number of data sets the hyper covariance matrix is often unidentifiable, a practical remedy is suggested through the eigenbasis transformation of the covariance matrix, which effectively reduces the number of unknown hyper-parameters. It is also proved that under some conditions the maximum a posteriori (MAP) estimation of the hyper mean and covariance coincide with the ensemble mean and covariance computed using the optimal estimations corresponding to multiple data sets. This interesting finding addresses relevant concerns related to the outcome of the mainstream Bayesian methods in capturing the stochastic variability from dissimilar data sets. Finally, the dynamical response of a prototype structure tested on a shaking table subjected to Gaussian white noise base excitation and the ambient vibration measurement of a cable footbridge are employed to demonstrate the proposed framework.

[1] PhD Candidate, Department of Civil and Environmental Engineering, The Hong Kong University of Science and Technology, Hong Kong, China; Department of Civil Engineering, Sharif University of Technology, Tehran, Iran, **Email:** sedehi.omid@gmail.com

[2] Professor, Department of Civil and Environmental Engineering, The Hong Kong University of Science and Technology, Hong Kong, China, **Email:** katafygiotis.lambros@gmail.com.

[3*] Professor, Department of Mechanical Engineering, University of Thessaly, Volos, Greece, **Email:** costasp@uth.gr (**Corresponding Author**)

# 1. Introduction

Modal identification of structural systems has been a crucial front in Structural Health Monitoring (SHM) [1]. In this respect, deterministic methods are commonly criticized mainly due to the presence of multiple sources of uncertainty, and this has shifted the attention toward using probabilistic methods [2–5]. These methods embed deterministic models within a class of prediction error probability models to describe the uncertainty in terms of the misfit between the system and model outputs. Two major perspectives are well recognized, namely the Frequentist and Bayesian approaches. The Maximum Likelihood Estimation (MLE) method is a well-known Frequentist approach, which has a relatively long history of usage in modal identification [6–10]. The MLE have similarities and connections with the Bayesian approach, especially in using probability axioms [11]. However, they are fundamentally different in the way they interpret the notion of probability [12]. The Bayesian approach characterizes the probability as the relative plausibility of a particular hypothesis/event within a set of candidates using fundamental probability axioms and conditional on the data [12,13]. This definition of probability is more flexible for quantifying the model and parameter uncertainties when compared with the Frequentist interpretation [12,14,15].

Yuen and Katafygiotis [16–19] have originally developed Bayesian modal identification methods in both time/frequency-domain using input-output or output-only data. Power Spectral Density (PSD) [17] and Fast Fourier Transform (FFT) [19] of the measured responses have been two common grounds for characterizing the dynamical models in the frequency-domain. However, model identifiability and computational burdens have overshadowed the applications of the original formulations until Au [20] has effectively addressed such implementation concerns and proposed the fast Bayesian FFT Operational Modal Analysis (OMA) method [21,22]. Substantial research effort has next been devoted to formulate the uncertainty laws [23,24], to combine multiple setup data [25–27], and to develop it further for handling buried modes, free vibrations, seismic excitations, as well as asynchronous data [28–31]. This method has also brought about successful applications in modal identification of real-life structures [22,32]. Motivated by these advances established on grounds of the FFT Bayesian approach, Yan and Katafygiotis [33,34] have elaborated on the computational issues of the Bayesian PSD formulations. This method is then developed for dealing with multiple sensor configurations [35], and its propagating properties are analytically studied in [36]. Using the PSD approach is advantageous for ambient modal identification since it separates the identification of the spectrum parameters (modal frequency and damping) from the spatial parameters (dynamical modes). Merits and demerits of the Bayesian FFT and PSD methods are well discussed in [37]. Although both the FFT and PSD Bayesian methods share similarities in terms of the concept, formulations, and applications, they differ in significant directions, including the averaging over multiple data segments required in the PSD approach. This feature of the PSD method is advantageous due to the promise it holds up for capturing the variability of system operating conditions across multiple data segments [34]. Nevertheless, the PSD method is suspected to be inaccurate for short-length response measurements mainly due to the invalidity asymptotic approximations [35,37]. Time-domain Bayesian modal identification methods are of the latest development in this respect [38–40].

It is believed that when little or no modeling error exists, Bayesian methods legitimately quantify the uncertainties [11]. In practice, however, this is not the case since the underlying deterministic and probability models are exposed to substantial inaccuracies. For instance, recent studies have highlighted that the dynamical parameters are commonly correlated with environmental parameters, ambient test conditions, vibration amplitude, input parameters, and other operating system conditions [41,42]. At the same time, the state-of-the-art dynamical models do not physically explain such factors since in practice they are extremely hard to control and determine, if not impossible. Consequently, the identification results noticeably vary over different experiments (data sets). This issue has led to criticism about the outcome of the mainstream Bayesian methods when this variability significantly exceeds the Bayesian estimation of the uncertainty. It is should be noted that such a comparison might not be rational since such Bayesian interpretation of the uncertainty solely characterizes the identification precision and is different from the variability due to modeling errors [11,22]. Consequently, accommodating the variability (among data uncertainty) and identification precision (within data uncertainty) into the posterior distributions through a robust, reliable, and rational framework has been a long-standing problem, being of both academic and practical significance [22].

Another conceptual limitation of Bayesian methods lies in treating the parameters as fixed unknown ones and regarding their uncertainty as reducible when additional observations are acquired. However, this is an inevitable prospect that the unknown parameters can also be exposed to some variability originating from modeling errors, which cannot be considered by the present Bayesian methods. These notable issues have discouraged investigators to use the Bayesian approach as a means for quantifying the uncertainties and resort to using an MLE approach instead. Such an approach has been associated with the Frequentist approach for characterizing the second-moment statistics and using the ensemble mean and covariance matrix computed from multiple experiments. As plainly stated in [12], nevertheless, using a Frequentist approach is undesirable due to limitations in justifying the model and parameter uncertainties, as well as its reliance on the concept of true value(s) of the parameters which might not exist at all.

Hierarchical Bayesian methods open up new horizons in modeling the uncertainty and have shown great promise in different scientific disciplines [43–46]. In structural dynamics, Behmanesh et al. [47] have established a hierarchy of parameters to capture the inherent variability of parameters while using experimental modal data. Recently, this framework has been used to quantify and propagate the uncertainty due to modeling errors, environmental variability, and excitation amplitude [48–51]. Nagel and Sudret [52,53] have developed a hierarchical Bayesian framework for the particular case of having noise-free vibration data. Sedehi et al. [54–56] have developed a hierarchical Bayesian framework for time-domain model updating and response predictions. When this framework is compared with the non-hierarchical Bayesian framework developed by Beck and Katafygiotis [15], it greatly outperforms the non-hierarchical framework in terms of the reliability and robustness of the uncertainty bounds computed for both the model parameters and response quantities of interest. The computation of this framework is built upon the existing Laplace asymptotic approximation methods. This allows exploiting the state-of-the-art Bayesian formulations under a hierarchical setting aiming to capture the variability prompted due to modeling errors. Such developments have been absent from the modal identification literature, and this study is a pioneering effort to pursue the same line of research as [56] and to develop it further in terms of the theory, computation, and applications, particularly for probabilistic modal identification problems. For this purpose, the fast Bayesian FFT formulations are used for ambient modal identification. Note that the Bayesian FFT approach suits better in the hierarchical framework in the sense that it does not need averaging over time-history segments and can infer the modal parameters directly from each data set. The computation of the hierarchical model is addressed using MCMC sampling and Laplace asymptotic approximation methods. Subsequently, two experimental examples are employed to examine and demonstrate the proposed method using vibration data.

The present study contributes to the development of the next generation of Bayesian structural identification methods that establish a more holistic framework for handling the challenging task of uncertainty quantification. It also provides insights on the uncertainty quantification in modal identification problems showing to what extent the variability dictated by the modeling errors could be dominant and how this hierarchical framework successfully handles it.

This paper continues with Section 2 reviewing the fast Bayesian FFT method. Section 3 describes the proposed hierarchical Bayesian framework, the graphical model, and the mathematical formulations. Section 4 discusses the computational aspects of the hierarchical model and develops two algorithms for this purpose. The dynamical response of a prototype structure tested on a shaking-table, as well as ambient vibration measurements of a cable footbridge will be employed in Section 5 to demonstrate the proposed framework. Conclusions and future works are drawn in the end.

## 2. Bayesian FFT-based modal inference

This section quickly reviews the fast Bayesian FFT modal inference method developed for well-separated modes, originally established in [19] and greatly enhanced later in [20,22]. The notation used to present the method is chosen to be the same as [22] for ease of reading and comparing with the recent studies. To begin with, let $D = \{\hat{\mathbf{y}}_j \in \mathbb{R}^n, j = 0,..., N-1\}$ denote a data set comprising discrete-time response of a dynamical system measured at $n$ degrees-of-freedom (DOF), where the index $j$ corresponds to the discrete-time $t_j = j\Delta t$ and $\Delta t$ is the sampling interval. Note that the response can be displacement, velocity, and acceleration measurements. The scaled FFT response is computed as

$$\hat{\mathbf{F}}_k = \sqrt{\frac{\Delta t}{N}} \sum_{j=0}^{N-1} \hat{\mathbf{y}}_j \exp\left(-\frac{2\pi \mathbf{i} jk}{N}\right) \qquad (1)$$

where $\mathbf{i}$ is the complex index; $\hat{\mathbf{F}}_k \in \mathbb{C}^n$ is the Fourier transform of the response corresponding to the frequency $f_k = k/(N\Delta t)$, $k = 1, 2, ..., N_q$. Note that the FFT response can be computed for frequencies smaller than the Nyquist frequency, which is equivalent to $k < N_q = \text{int}[N/2]$, where int[.] denotes the integer part of a real number.

To construct a parametric model in frequency-domain, the response of a linear time-invariant dynamical system is considered. Let $f_k \in [f_{lb}, f_{ub}]$ represent a frequency band containing $N_f$ samples selected such that only the resonance peak corresponding to the $i^{th}$ dynamical mode appears on the so-called singular value spectrum. Due to the Newton's second law of motion, the Fourier transform of the response within this particular band transferred into the modal coordinates can be expressed as

$$\mathbf{F}_k = \boldsymbol{\varphi}_i h_{ik} p_{ik} / \|\boldsymbol{\varphi}_i\| \qquad (2)$$

where the subscript $k$ represents the frequency $f_k$, $\boldsymbol{\varphi}_i \in \mathbb{R}^n$ is the $i^{th}$ dynamical mode shape, $\|.\|$ denotes the Euclidean norm of a vector, $p_{ik}$ is the scaled FFT of the $i^{th}$ modal force, and $h_{ik}$ is the frequency response function determined from

$$h_{ik} = \frac{(2\pi \mathbf{i} f_k)^{-q}}{1 - \beta_{ik}^2 - 2\xi_i \beta_{ik} \mathbf{i}} \qquad (3)$$

Here, $q$ takes on 0, 1, and 2 for acceleration, velocity, and displacement response measurements, respectively; $\beta_{ik}$ is the frequency ratio $f_k / f_i$; $f_i$ and $\xi_i$ are the modal frequency and damping ratio corresponding to the $i^{th}$ dynamical mode. This parametric model predicts the actual FFT response as

$$\hat{\mathbf{F}}_k = \mathbf{F}_k + \boldsymbol{\varepsilon}_k \qquad (4)$$

where $\boldsymbol{\varepsilon}_k \in \mathbb{R}^n$ denotes the prediction error reflecting the mismatch between the actual and the model responses in the frequency domain. Let the prediction errors be statistically independent and identically distributed (*i.i.d.*) described using Gaussian distributions, i.e.

$$\boldsymbol{\varepsilon}_k \underset{i.i.d.}{\sim} N(\mathbf{0}, S_{e,i} \mathbf{I}_n) \tag{5}$$

where $S_{e,i}$ is prediction error variance corresponding to the $i^{\text{th}}$ dynamical mode, and $\mathbf{I}_n \in \mathbb{R}^{n \times n}$ is the identity matrix. This variance is considered to be fixed over the entire band of interest and across all DOF.

Recall that $N_f$ denotes the number of FFT response samples existing within the frequency band of interest. For a large number of data points when $N_f \gg 1$, the FFT response can be described using a complex Gaussian distribution expressed as [22]

$$p(\hat{\mathbf{F}}_k | \boldsymbol{\theta}) = \pi^{-n} |\mathbf{E}_k(\boldsymbol{\theta})|^{-1} \exp\left(-\hat{\mathbf{F}}_k^* \mathbf{E}_k^{-1}(\boldsymbol{\theta}) \hat{\mathbf{F}}_k\right) \tag{6}$$

where $\boldsymbol{\theta} = \begin{bmatrix} f_i & \xi_i & \boldsymbol{\varphi}_i^T & S_i & S_{ei} \end{bmatrix}^T$ denotes the model parameters, $\mathbf{E}_k(\boldsymbol{\theta}) = S_i D_{ik} \boldsymbol{\varphi}_i \boldsymbol{\varphi}_i^T / \|\boldsymbol{\varphi}_i\|^2 + S_{ei} \mathbf{I}_n$ is the theoretical PSD matrix parameterized by $\boldsymbol{\theta}$, $S_i$ is the PSD of the modal force $p_{ik}$ assumed to constant over the selected band, and $D_{ik} = h_{ik} h_{ik}^*$ is the dynamic amplification factor. Considering $\hat{\mathbf{F}}_k$'s as statistically independent realizations gives the likelihood function as

$$p(\{\hat{\mathbf{F}}_k\} | \boldsymbol{\theta}) = \prod_k p(\hat{\mathbf{F}}_k | \boldsymbol{\theta}) \tag{7}$$

where $\{\hat{\mathbf{F}}_k\}$ comprises all frequency response components fallen within the chosen band. The Bayes' rule gives the joint posterior distribution as

$$p(\boldsymbol{\theta} | \{\hat{\mathbf{F}}_k\}) \propto p(\{\hat{\mathbf{F}}_k\} | \boldsymbol{\theta}) p(\boldsymbol{\theta}) \tag{8}$$

where $p(\boldsymbol{\theta})$ is prior distribution described using uniform distributions, and $p(\boldsymbol{\theta} | \{\hat{\mathbf{F}}_k\})$ is posterior distribution, which regularly appears to be "locally identifiable" [57] as it draws a sharp peak in a specific neighborhood representing the Most Probable Values (MPV). Computing the MPV requires

the maximization of the posterior distribution with respect to the model parameters. However, it is computationally desirable to minimize the negative logarithm of the posterior distribution given as

$$L(\boldsymbol{\theta}) = nN_f \ln \pi + \sum_k \ln |\mathbf{E}_k(\boldsymbol{\theta})| + \sum_k \hat{\mathbf{F}}_k^* \mathbf{E}_k^{-1}(\boldsymbol{\theta}) \hat{\mathbf{F}}_k + c \qquad (9)$$

Note that these summations are over the $N_f$ data points given within the selected band and $c$ is a constant value. Having considered this objective function, the MPV ($\hat{\boldsymbol{\theta}}$) is determined from

$$\hat{\boldsymbol{\theta}} = \underset{\boldsymbol{\theta}}{\text{Argmin}}\ L(\boldsymbol{\theta}) \qquad (10)$$

As the total number of data points ($nN_f$) grows, the posterior distribution concentrates around the MPV. Consequently, for a sufficiently large number of data points, a Laplace asymptotic approximation provides an efficient approximation of the posterior distribution giving [14,15,58]:

$$p(\boldsymbol{\theta} | \{\hat{\mathbf{F}}_k\}) \approx N(\boldsymbol{\theta} | \hat{\boldsymbol{\theta}}, \hat{\boldsymbol{\Sigma}}_{\boldsymbol{\theta\theta}}) \qquad (11)$$

where $\hat{\boldsymbol{\Sigma}}_{\boldsymbol{\theta\theta}} = [\nabla_{\boldsymbol{\theta}} \nabla_{\boldsymbol{\theta}}^T L(\boldsymbol{\theta})]^{-1}_{\boldsymbol{\theta} = \hat{\boldsymbol{\theta}}}$ is the covariance matrix determined as the inverse of the Hessian matrix of $L(\boldsymbol{\theta})$ evaluated at the MPV ($\hat{\boldsymbol{\theta}}$). It should be pointed out that this optimization problem can reliably be carried out when one supplies reasonable initial estimations and explicit derivatives of the objective function. This issue is addressed in the literature, and the readers can refer to [22] for a reliable computation of $\hat{\boldsymbol{\theta}}$ and $\hat{\boldsymbol{\Sigma}}_{\boldsymbol{\theta\theta}}$.

**Remark-1**. Given that the prior distribution is uniform, the likelihood function and the posterior distribution are proportional. Therefore, the Laplace approximation in Eq. (11) is also valid to simplify the likelihood function. This feature is the key for integrating the Fast Bayesian FFT approach within the hierarchical framework appearing in the next section.

**Remark-2**. In the above formulations, the mode shape vector is considered to have unit Euclidean norm. Thus, when the derivatives of the objective function $L(\boldsymbol{\theta})$ are computed, the norm constraints are explicitly considered like in [22].

## 3. Hierarchical Bayesian modeling framework

### 3.1. Probabilistic model

In this section, we introduce a novel hierarchical modeling framework to consider the variability of parameters over multiple experiments. Let $\mathbf{D} = \{D_s, s = 1, ..., N_D\}$ be a family of data sets comprising $N_D$ independent sets of vibration measurements, where $D_s = \{\hat{\mathbf{y}}_{j,s} \in \mathbb{R}^n, j = 0, ..., N_s - 1\}$ is a particular data set comprising time-history responses. Note that in the remainder wherever the subscript $s$ and $j$ appear, they respectively correspond to the data set $D_s$ and the time instant $t_j = j\Delta t$. For simplicity and notational convenience, the sampling interval of different data sets is considered to be equal, although in general they can be different. Each data set ($D_s$) is assumed to have a sufficiently large number of data points, from which the model parameters can be inferred. Satisfying this condition is not difficult in practice, especially when the time-history measurements are available from a permanent sensor network.

The likelihood function of each data set can be constructed in the frequency-domain similar to the procedure presented through Eqs. (1-7). Moreover, due to Remark-1, the Laplace approximation can be used to simplify the likelihood function giving:

$$p(D_s | \boldsymbol{\theta}_s) \propto N(\boldsymbol{\theta}_s | \hat{\boldsymbol{\theta}}_s, \hat{\boldsymbol{\Sigma}}_{\boldsymbol{\theta}_s \boldsymbol{\theta}_s}) \tag{12}$$

where $p(D_s | \boldsymbol{\theta}_s)$ is the likelihood function corresponding to $D_s$; $\boldsymbol{\theta}_s = [f_{i,s} \; \xi_{i,s} \; \boldsymbol{\varphi}_{i,s}^T \; S_{i,s} \; S_{ei,s}]^T$ denotes the model parameters corresponding to data set $D_s$; $\hat{\boldsymbol{\theta}}_s$ and $\hat{\boldsymbol{\Sigma}}_{\boldsymbol{\theta}_s \boldsymbol{\theta}_s}$ are respectively the mean and covariance computed based on Remark-1 and using the Fast Bayesian FFT approach [22] wherein uniform prior distribution is used. Note that adding the subscript $s$ aims to underline that the parameters inferred from each data set are unequal and variable. It should also be emphasized that Eq. (12) holds in the proportional sense and only approximately.

We introduce a specific re-parameterization to collect the dynamical parameters in $\boldsymbol{\lambda}_s \triangleq [f_{i,s}, \xi_{i,s}, \boldsymbol{\varphi}_{i,s}^T]^T \in \mathbb{R}^{n+2}$ and the non-dynamical parameters in $\boldsymbol{\eta}_s \triangleq [S_{i,s}, S_{ei,s}]^T \in \mathbb{R}^2$. The dynamical parameters are considered to have common statistical properties over multiple data sets, whereas the non-dynamical parameters can vary across data sets as statistically independent random variables. In contrast, we consider the dynamical parameters, $\boldsymbol{\lambda}_s$'s, as the randomly-drawn realizations of a hyper probability distribution, which encapsulates the statistical information of the dynamical parameters ($\boldsymbol{\lambda}_s$). The hyper probability distribution is described through a multivariate Gaussian distribution expressed as $N(\boldsymbol{\lambda}_s | \boldsymbol{\mu}_\lambda, \boldsymbol{\Sigma}_{\lambda\lambda})$, where $\boldsymbol{\mu}_\lambda \in \mathbb{R}^{n+2}$ and $\boldsymbol{\Sigma}_{\lambda\lambda} \in \mathbb{R}^{(n+2)\times(n+2)}$ are the uncertain mean and covariance, respectively. Choosing Gaussian distribution offers optimality in the maximum entropy sense [13]. We represent this hyper distribution by $P(\psi)$, where $\psi$ is the hyper-parameter set $\psi = \{\boldsymbol{\mu}_\lambda, \boldsymbol{\Sigma}_{\lambda\lambda}\}$ whose elements are reparametrized as

$$\boldsymbol{\mu}_\lambda = \begin{bmatrix} \mu_{f_i} \\ \mu_{\xi_i} \\ \boldsymbol{\mu}_{\boldsymbol{\varphi}_i} \end{bmatrix}_{(n+2)\times 1} ; \quad \boldsymbol{\Sigma}_{\lambda\lambda} = \begin{bmatrix} \sigma_{f_i}^2 & \sigma_{f_i \xi_i}^2 & \boldsymbol{\Sigma}_{f_i \boldsymbol{\varphi}_i} \\ \sigma_{f_i \xi_i}^2 & \sigma_{\xi_i}^2 & \boldsymbol{\Sigma}_{\xi_i \boldsymbol{\varphi}_i} \\ \boldsymbol{\Sigma}_{f_i \boldsymbol{\varphi}_i}^T & \boldsymbol{\Sigma}_{\xi_i \boldsymbol{\varphi}_i}^T & \boldsymbol{\Sigma}_{\boldsymbol{\varphi}_i \boldsymbol{\varphi}_i} \end{bmatrix}_{(n+2)\times(n+2)} \quad (13)$$

where the elements with subscripts $f_i$, $\xi_i$ and $\boldsymbol{\varphi}_i$ correspond to the modal frequency, damping ratio, and mode shape vector, respectively.

Fig. 1 shows the proposed hierarchical probabilistic model represented using plate notations. As shown, the hyper prior distribution will govern the variation of dynamical parameters ($\boldsymbol{\lambda}_s$) across $N_D$ data sets. There are $N_f$ samples of the response at the chosen resonant band represented by $\hat{\mathbf{F}}_k$. The model response ($\mathbf{F}_k$) is expressed as a function of dynamical parameters ($\boldsymbol{\lambda}_s$) and the PSD of modal force ($S_s$). The predictions errors are considered to be Gaussian white noise (GWN) processes with fixed PSD ($S_{e,s}$). Note that all these parameters correspond to a specific dynamical mode of interest.

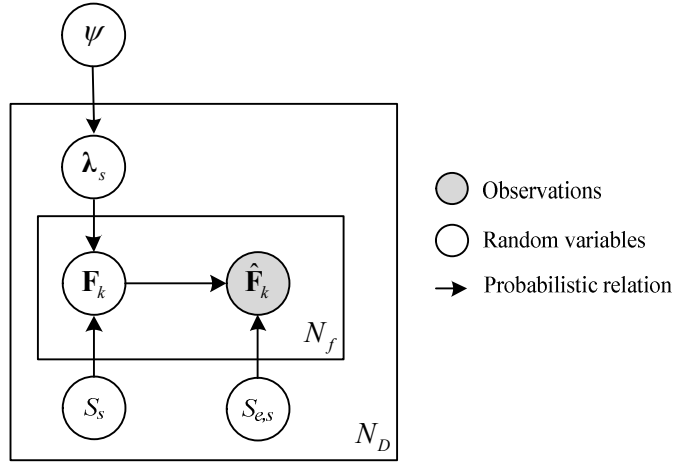

**Fig. 1.** Acyclic graphical representation of the proposed probabilistic model (The largest rectangular represents different data sets, while the smaller one shows the data points within each data set; $\psi = \{\boldsymbol{\mu}_\lambda, \boldsymbol{\Sigma}_{\lambda\lambda}\}$ is the hyper-parameters set)

### 3.2. Bayesian inference

*3.2.1. Joint posterior distribution*

In this section, we present how the proposed model with its hierarchical architecture can be updated and calibrated using multiple data sets. Given the model architecture shown in Fig. 1, the joint prior distribution is expressed as

$$p\left(\{\boldsymbol{\lambda}_s,\boldsymbol{\eta}_s\}_{s=1}^{N_D},\boldsymbol{\mu}_\lambda,\boldsymbol{\Sigma}_{\lambda\lambda}\right) \\ = p\left(\{\boldsymbol{\lambda}_s,\boldsymbol{\eta}_s\}_{s=1}^{N_D} \mid \boldsymbol{\mu}_\lambda,\boldsymbol{\Sigma}_{\lambda\lambda}\right)p\left(\boldsymbol{\mu}_\lambda,\boldsymbol{\Sigma}_{\lambda\lambda}\right) = p\left(\boldsymbol{\mu}_\lambda,\boldsymbol{\Sigma}_{\lambda\lambda}\right)\prod_{s=1}^{N_D} p\left(\boldsymbol{\lambda}_s \mid \boldsymbol{\mu}_\lambda,\boldsymbol{\Sigma}_{\lambda\lambda}\right)p\left(\boldsymbol{\eta}_s\right) \qquad (14)$$

where $p(\boldsymbol{\mu}_\lambda,\boldsymbol{\Sigma}_{\lambda\lambda})$ is prior distribution of the hyper-parameters, $p(\boldsymbol{\eta}_s)$ is prior distribution of the non-dynamical parameters, and the conditional distribution $p(\boldsymbol{\lambda}_s \mid \boldsymbol{\mu}_\lambda,\boldsymbol{\Sigma}_{\lambda\lambda})$ is specified earlier as $N(\boldsymbol{\lambda}_s \mid \boldsymbol{\mu}_\lambda,\boldsymbol{\Sigma}_{\lambda\lambda})$. Notice that this form of the prior distribution is obtained using the independence of $\boldsymbol{\lambda}_s$'s and $\boldsymbol{\eta}_s$'s across different data sets. Subsequently, this prior distribution can be updated based on the Bayes' rule and using the family of data sets, $\mathbf{D} = \{D_s\}_{s=1}^{N_D}$, which yields:

$$p\left(\{\boldsymbol{\lambda}_s,\boldsymbol{\eta}_s\}_{s=1}^{N_D},\boldsymbol{\mu}_\lambda,\boldsymbol{\Sigma}_{\lambda\lambda} \mid \{D_s\}_{s=1}^{N_D}\right) \propto p\left(\{D_s\}_{s=1}^{N_D} \mid \{\boldsymbol{\lambda}_s,\boldsymbol{\eta}_s\}_{s=1}^{N_D},\boldsymbol{\mu}_\lambda,\boldsymbol{\Sigma}_{\lambda\lambda}\right) p\left(\{\boldsymbol{\lambda}_s,\boldsymbol{\eta}_s\}_{s=1}^{N_D},\boldsymbol{\mu}_\lambda,\boldsymbol{\Sigma}_{\lambda\lambda}\right) \qquad (15)$$

Here, $p\left(\{D_s\}_{s=1}^{N_D} \mid \{\boldsymbol{\lambda}_s, \boldsymbol{\eta}_s\}_{s=1}^{N_D}, \boldsymbol{\mu}_\lambda, \boldsymbol{\Sigma}_{\lambda\lambda}\right)$ is the likelihood function of all data sets. Owing to the statistical independence of the data sets, the likelihood of the family of data sets can expressed as

$$p\left(\{D_s\}_{s=1}^{N_D} \mid \{\boldsymbol{\lambda}_s, \boldsymbol{\eta}_s\}_{s=1}^{N_D}, \boldsymbol{\mu}_\lambda, \boldsymbol{\Sigma}_{\lambda\lambda}\right) = \prod_{s=1}^{N_D} p\left(D_s \mid \boldsymbol{\lambda}_s, \boldsymbol{\eta}_s\right) \tag{16}$$

where the likelihood function $p(D_s \mid \boldsymbol{\lambda}_s, \boldsymbol{\eta}_s)$ can be approximated based on Eq. (12) giving:

$$p(D_s \mid \boldsymbol{\lambda}_s, \boldsymbol{\eta}_s) \propto N\left(\begin{bmatrix} \boldsymbol{\lambda}_s \\ \boldsymbol{\eta}_s \end{bmatrix} \middle| \begin{bmatrix} \hat{\boldsymbol{\lambda}}_s \\ \hat{\boldsymbol{\eta}}_s \end{bmatrix}, \begin{bmatrix} \hat{\boldsymbol{\Sigma}}_{\lambda_s \lambda_s} & \hat{\boldsymbol{\Sigma}}_{\lambda_s \eta_s} \\ \hat{\boldsymbol{\Sigma}}_{\lambda_s \eta_s}^T & \hat{\boldsymbol{\Sigma}}_{\eta_s \eta_s} \end{bmatrix}\right) \tag{17}$$

In this equation, $\hat{\boldsymbol{\lambda}}_s$ and $\hat{\boldsymbol{\eta}}_s$ are the corresponding elements from the mean vector $\hat{\boldsymbol{\theta}}_s$; $\hat{\boldsymbol{\Sigma}}_{\lambda_s \lambda_s}$, $\hat{\boldsymbol{\Sigma}}_{\lambda_s \eta_s}$, and $\hat{\boldsymbol{\Sigma}}_{\eta_s \eta_s}$ are the corresponding blocks from the covariance matrix $\hat{\boldsymbol{\Sigma}}_{\theta_s \theta_s}$. Notice that in Eq. (16) the likelihood of the family of data sets does not depend on the hyper-parameters due to the particular choice of the probability model structure. Combining Eqs. (14-17) gives the joint posterior distribution as

$$\begin{aligned} & p\left(\{\boldsymbol{\lambda}_s, \boldsymbol{\eta}_s\}_{s=1}^{N_D}, \boldsymbol{\mu}_\lambda, \boldsymbol{\Sigma}_{\lambda\lambda} \mid \{D_s\}_{s=1}^{N_D}\right) \\ & \propto p(\boldsymbol{\mu}_\lambda, \boldsymbol{\Sigma}_{\lambda\lambda}) \prod_{s=1}^{N_D} \left[ N\left(\begin{bmatrix} \boldsymbol{\lambda}_s \\ \boldsymbol{\eta}_s \end{bmatrix} \middle| \begin{bmatrix} \hat{\boldsymbol{\lambda}}_s \\ \hat{\boldsymbol{\eta}}_s \end{bmatrix}, \begin{bmatrix} \hat{\boldsymbol{\Sigma}}_{\lambda_s \lambda_s} & \hat{\boldsymbol{\Sigma}}_{\lambda_s \eta_s} \\ \hat{\boldsymbol{\Sigma}}_{\lambda_s \eta_s}^T & \hat{\boldsymbol{\Sigma}}_{\eta_s \eta_s} \end{bmatrix}\right) N(\boldsymbol{\lambda}_s \mid \boldsymbol{\mu}_\lambda, \boldsymbol{\Sigma}_{\lambda\lambda}) p(\boldsymbol{\eta}_s) \right] \end{aligned} \tag{18}$$

*3.2.2. Marginal distributions*

As the hyper-parameters encapsulate the statistical information of the dynamical properties, it is desirable to compute its marginal posterior distribution. This task involves marginalizing $\boldsymbol{\lambda}_s$'s and $\boldsymbol{\eta}_s$'s from the joint distribution. Fortunately, the integrand is separable with respect to the dynamical and non-dynamical parameters. We first integrate the joint distribution over $\boldsymbol{\eta}_s$'s. When $p(\boldsymbol{\eta}_s)$ is considered to be uniform over the parameters' space, the integrand will only be the Gaussian approximation of the likelihood function. Then, integrating $\boldsymbol{\eta}_s$ becomes straightforward and only requires selecting those entries corresponding to the dynamical parameters, which leads to the following explicit formulation:

$$p\left(\{\boldsymbol{\lambda}_s\}_{s=1}^{N_D}, \boldsymbol{\mu}_\lambda, \boldsymbol{\Sigma}_{\lambda\lambda} \mid \{D_s\}_{s=1}^{N_D}\right) \propto p(\boldsymbol{\mu}_\lambda, \boldsymbol{\Sigma}_{\lambda\lambda}) \prod_{s=1}^{N_D} \left[ N\left(\boldsymbol{\lambda}_s \mid \hat{\boldsymbol{\lambda}}_s, \hat{\boldsymbol{\Sigma}}_{\lambda_s \lambda_s}\right) N(\boldsymbol{\lambda}_s \mid \boldsymbol{\mu}_\lambda, \boldsymbol{\Sigma}_{\lambda\lambda}) \right] \quad (19)$$

Note that to derive this equation we benefitted from a basic property of Gaussian distributions, that is, the marginal distribution of every selected subset of the parameters jointly described by a multivariate Gaussian distribution is also Gaussian whose mean and covariance can be determined by selecting the corresponding elements from the mean and covariance matrix of the joint distribution [59].

After marginalizing the non-dynamical parameters, the next step is to carry out the integration over the data-set-specific dynamical parameters. This integration involves the two Gaussian distributions inside the multiplication in Eq. (19), which is proved to have the following analytical solution (Appendix A):

$$\begin{aligned} p\left(\boldsymbol{\mu}_\lambda, \boldsymbol{\Sigma}_{\lambda\lambda} \mid \{D_s\}_{s=1}^{N_D}\right) &\propto p(\boldsymbol{\mu}_\lambda, \boldsymbol{\Sigma}_{\lambda\lambda}) \prod_{s=1}^{N_D} \left[ \int_{\boldsymbol{\lambda}_s} N\left(\boldsymbol{\lambda}_s \mid \hat{\boldsymbol{\lambda}}_s, \hat{\boldsymbol{\Sigma}}_{\lambda_s \lambda_s}\right) N(\boldsymbol{\lambda}_s \mid \boldsymbol{\mu}_\lambda, \boldsymbol{\Sigma}_{\lambda\lambda}) d\boldsymbol{\lambda}_s \right] \\ &\propto p(\boldsymbol{\mu}_\lambda, \boldsymbol{\Sigma}_{\lambda\lambda}) \prod_{s=1}^{N_D} N\left(\boldsymbol{\mu}_\lambda \mid \hat{\boldsymbol{\lambda}}_s, \boldsymbol{\Sigma}_{\lambda\lambda} + \hat{\boldsymbol{\Sigma}}_{\lambda_s \lambda_s}\right) \end{aligned} \quad (20)$$

where $p\left(\boldsymbol{\mu}_\lambda, \boldsymbol{\Sigma}_{\lambda\lambda} \mid \{D_s\}_{s=1}^{N_D}\right)$ is the marginal posterior distribution of the hyper-parameters. This derivation is central in the proposed hierarchical Bayesian framework since it will later be used for computing the hyper-parameters MAP estimation. Moreover, it describes both the variability and identification precision of the dynamical parameters.

It is useful to compute the marginal posterior distribution of the data-set-specific dynamical parameters. This requires fusing the information from multiple data sets to infer the dynamical parameters of each data set. This task can be carried out under the proposed hierarchical setting by computing the marginal distribution $p\left(\boldsymbol{\lambda}_r \mid \{D_s\}_{s=1}^{N_D}\right)$ from Eq. (19), where $r = \{1, 2, ..., N_D\}$. In pursuit of this goal, like Eq. (20) when all $\boldsymbol{\lambda}_s$'s except $\boldsymbol{\lambda}_r$ are marginalized out from Eq. (19), one arrives at

$$p\left(\boldsymbol{\lambda}_r, \boldsymbol{\mu}_\lambda, \boldsymbol{\Sigma}_{\lambda\lambda} \mid \{D_s\}_{s=1}^{N_D}\right) \propto N\left(\boldsymbol{\lambda}_r \mid \hat{\boldsymbol{\lambda}}_r, \hat{\boldsymbol{\Sigma}}_{\lambda_r \lambda_r}\right) N(\boldsymbol{\lambda}_r \mid \boldsymbol{\mu}_\lambda, \boldsymbol{\Sigma}_{\lambda\lambda}) p\left(\boldsymbol{\mu}_\lambda, \boldsymbol{\Sigma}_{\lambda\lambda} \mid \{D_s\}_{s=1, s \neq r}^{N_D}\right) \quad (21)$$

where the hyper-parameters distribution is expressed as

$$p\left(\boldsymbol{\mu}_\lambda, \boldsymbol{\Sigma}_{\lambda\lambda} \mid \{D_s\}_{s=1,s\neq r}^{N_D}\right) \propto p\left(\boldsymbol{\mu}_\lambda, \boldsymbol{\Sigma}_{\lambda\lambda}\right) \prod_{s=1,s\neq r}^{N_D} \left[ N\left(\boldsymbol{\mu}_\lambda \mid \hat{\boldsymbol{\lambda}}_s, \boldsymbol{\Sigma}_{\lambda\lambda} + \hat{\boldsymbol{\Sigma}}_{\lambda_s\lambda_s}\right) \right] \quad (22)$$

It should be pointed out that the difference between the L.H.S. of Eqs. (20) and (22) can be subtle for a sufficiently large number of data sets. In this case, we can compute only the integration in Eq. (20) and use it as a substitution for $p\left(\boldsymbol{\mu}_\lambda, \boldsymbol{\Sigma}_{\lambda\lambda} \mid \{D_s\}_{s=1,s\neq r}^{N_D}\right)$.

As proved in Appendix (A), the multiplication of Gaussian distributions in Eq. (21) can be expressed as

$$N\left(\boldsymbol{\lambda}_r \mid \hat{\boldsymbol{\lambda}}_r, \hat{\boldsymbol{\Sigma}}_{\lambda_r\lambda_r}\right) N\left(\boldsymbol{\lambda}_r \mid \boldsymbol{\mu}_\lambda, \boldsymbol{\Sigma}_{\lambda\lambda}\right) \propto N\left(\boldsymbol{\lambda}_r \mid \hat{\boldsymbol{\lambda}}_r + \mathbf{K}_\lambda\left(\boldsymbol{\mu}_\lambda - \hat{\boldsymbol{\lambda}}_r\right), \hat{\boldsymbol{\Sigma}}_{\lambda_r\lambda_r} - \mathbf{K}_\lambda \hat{\boldsymbol{\Sigma}}_{\lambda_r\lambda_r}\right) \quad (23)$$

where $\mathbf{K}_\lambda = \hat{\boldsymbol{\Sigma}}_{\lambda_r\lambda_r} \left(\hat{\boldsymbol{\Sigma}}_{\lambda_r\lambda_r} + \boldsymbol{\Sigma}_{\lambda\lambda}\right)^{-1}$ is the so-called gain matrix [59,60]. Substituting this result into Eq. (21) reads as

$$p\left(\boldsymbol{\lambda}_r, \boldsymbol{\mu}_\lambda, \boldsymbol{\Sigma}_{\lambda\lambda} \mid \{D_s\}_{s=1}^{N_D}\right) \propto N\left(\boldsymbol{\lambda}_r \mid \hat{\boldsymbol{\lambda}}_r + \mathbf{K}_\lambda\left(\boldsymbol{\mu}_\lambda - \hat{\boldsymbol{\lambda}}_r\right), \hat{\boldsymbol{\Sigma}}_{\lambda_r\lambda_r} - \mathbf{K}_\lambda \hat{\boldsymbol{\Sigma}}_{\lambda_r\lambda_r}\right) p\left(\boldsymbol{\mu}_\lambda, \boldsymbol{\Sigma}_{\lambda\lambda} \mid \{D_s\}_{s=1,s\neq r}^{N_D}\right) \quad (24)$$

Marginalizing the hyper-parameters from this distribution leads to the desired distribution computed as

$$p\left(\boldsymbol{\lambda}_r \mid \{D_s\}_{s=1}^{N_D}\right) = \int_{\boldsymbol{\Sigma}_{\lambda\lambda}} \int_{\boldsymbol{\mu}_\lambda} N\left(\boldsymbol{\lambda}_r \mid \hat{\boldsymbol{\lambda}}_r + \mathbf{K}_\lambda\left(\boldsymbol{\mu}_\lambda - \hat{\boldsymbol{\lambda}}_r\right), \hat{\boldsymbol{\Sigma}}_{\lambda_r\lambda_r} - \mathbf{K}_\lambda \hat{\boldsymbol{\Sigma}}_{\lambda_r\lambda_r}\right) p\left(\boldsymbol{\mu}_\lambda, \boldsymbol{\Sigma}_{\lambda\lambda} \mid \{D_s\}_{s=1,s\neq r}^{N_D}\right) d\boldsymbol{\mu}_\lambda d\boldsymbol{\Sigma}_{\lambda\lambda} \quad (25)$$

The latest equation characterizes the identification precision corresponding to the $r^{th}$ data set/experiment computed using the information from the family of data sets.

### 3.2.3. Posterior predictive distribution

For unobserved operating conditions, the hyper-parameters' uncertainty can be propagated to compute the posterior predictive distribution of the dynamical parameters. Let $\boldsymbol{\lambda}_{N_D+1}$ denote the dynamical parameters of an unobserved operating conditions while $N_D$ data sets are obtained. As prior information, the random realization ($\boldsymbol{\lambda}_{N_D+1}$) is assumed to be drawn from the foregoing hyper distribution described as $p(\boldsymbol{\lambda}_{N_D+1}|\boldsymbol{\mu}_\lambda,\boldsymbol{\Sigma}_{\lambda\lambda}) \triangleq N(\boldsymbol{\lambda}_{N_D+1}|\boldsymbol{\mu}_\lambda,\boldsymbol{\Sigma}_{\lambda\lambda})$. Consequently, the uncertainty associated with this distribution given the data can be propagated to predict $\boldsymbol{\lambda}_{N_D+1}$ using the principle of total probability:

$$p(\boldsymbol{\lambda}_{N_D+1}|\{D_s\}_{s=1}^{N_D}) = \int_{\boldsymbol{\Sigma}_{\lambda\lambda}}\int_{\boldsymbol{\mu}_\lambda} N(\boldsymbol{\lambda}_{N_D+1}|\boldsymbol{\mu}_\lambda,\boldsymbol{\Sigma}_{\lambda\lambda}) p(\boldsymbol{\mu}_\lambda,\boldsymbol{\Sigma}_{\lambda\lambda}|\{D_s\}_{s=1}^{N_D}) d\boldsymbol{\mu}_\lambda d\boldsymbol{\Sigma}_{\lambda\lambda} \quad (26)$$

where $p(\boldsymbol{\lambda}_{N_D+1}|\{D_s\}_{s=1}^{N_D})$ is the posterior predictive distribution of dynamical parameters, which captures the variability of dynamical parameters over multiple experiments.

In summary, the quantification of the uncertainty associated with the hyper-parameters, as well as the data-set-specific dynamical parameters can be accomplished based on Eqs. (20), (25), and (26). However, integrations appearing in these equations are often accompanied by computational difficulties discussed in the next section.

## 4. Computational strategies

### 4.1. MCMC sampling

The multi-dimensional integrals in Eqs. (25-26) can be carried out using MCMC sampling methods. This needs to draw samples of the hyper-parameters from the probability distributions specified in either Eq. (20) or (22). We demonstrate the sampling method only for the former case as the latter can be accomplished similarly.

The structure of the hyper covariance matrix maps out our sampling strategy. When $\boldsymbol{\Sigma}_{\lambda\lambda}$ is diagonal, the MCMC sampling method is straightforward addressed in [47,61]. When the covariance matrices $\hat{\boldsymbol{\Sigma}}_{\lambda_s\lambda_s}$'s are also diagonal, the sampling over the hyper-parameters' marginal distribution is even more efficient. To clarify the latter case, let $\mu_{\lambda,p}$ and $\sigma^2_{\lambda,pp}$ be one pair of the parameters selected from the hyper mean vector and the diagonal hyper covariance matrix, where the subscript $p = \{1,2,...,n+2\}$ represents the $p^{\text{th}}$ pair of the parameters. Thus, the sampling can be performed over each pair of the hyper-parameters taken from the following distribution:

$$p\left(\mu_{\lambda,p},\sigma_{\lambda,pp}^2 \mid \{D_s\}_{s=1}^{N_D}\right) \propto p\left(\mu_{\lambda,p},\sigma_{\lambda,pp}^2\right) \prod_{s=1}^{N_D} N\left(\mu_{\lambda,p} \mid \hat{\lambda}_{s,p}, \sigma_{\lambda,pp}^2 + \hat{\sigma}_{\lambda_s\lambda_s,pp}^2\right) \qquad (27)$$

where $\hat{\lambda}_{s,pp}$ and $\hat{\sigma}_{\lambda_s\lambda_s,pp}^2$ are the corresponding elements from $\hat{\boldsymbol{\lambda}}_s$ and $\hat{\boldsymbol{\Sigma}}_{\lambda_s\lambda_s}$, respectively. However, this simplification is inapplicable for non-diagonal covariance matrices, and applying such a naive approach which directly attempts to sample the hyper-parameters would be infeasible mainly due to difficulties in retaining the positive semi-definiteness conditions. To resolve this problem, we first reparametrize the elements of the covariance matrix as $\Sigma_{\lambda\lambda,pq} = \rho_{\lambda\lambda,pq}\sigma_{\lambda\lambda,pp}\sigma_{\lambda\lambda,qq}$, where $\sigma_{\lambda\lambda,pp}$ is the standard deviation of $\lambda_p$; $\forall p,q = \{1,2,...,n\}$: $\rho_{\lambda\lambda,pq} = \rho_{\lambda\lambda,qp}$ & $\rho_{\lambda\lambda,pp} = 1$ represents the correlation between $\lambda_p$ and $\lambda_q$; $\lambda_p$ represents the $p^{\text{th}}$ element of $\boldsymbol{\lambda}$. This reparametrization enforces the symmetric structure of the hyper covariance matrix and requires the sampling to be performed only over the parameters $\sigma_{\lambda\lambda,pp}$ and $\rho_{\lambda\lambda,pq}$ rather than the full covariance matrix. Moreover, the positive semi-definiteness of the samples of the covariance matrix can be preserved using the Cholesky decomposition expressing it as $\boldsymbol{\Sigma}_{\lambda\lambda} = \mathbf{S}_{\lambda\lambda}\mathbf{L}_{\lambda\lambda}\mathbf{L}_{\lambda\lambda}^T\mathbf{S}_{\lambda\lambda}$, where $\mathbf{L}_{\lambda\lambda} \in \mathbb{R}^{(n+2)\times(n+2)}$ is a lower-triangular unitary matrix and $\mathbf{S}_{\lambda\lambda} \in \mathbb{R}^{(n+2)\times(n+2)}$ is a diagonal matrix comprising the standard deviations $\sigma_{\lambda\lambda,pp}$. Analytical formulations for computing the Cholesky decomposition in terms of the correlation coefficients and the variances are derived in [62] and briefly described in Appendix (B). This decomposition accompanied by the conditions $\sigma_{\lambda\lambda,mm} \geq 0$ and $-1 < \rho_{\lambda\lambda,mn} < 1$ assist to retain the symmetry and positive semi-definiteness conditions.

Once the MCMC samples of the hyper-parameters are obtained, the integral in Eq. (25) can be approximated as

$$p\left(\boldsymbol{\lambda}_r \mid \{D_s\}_{s=1}^{N_D}\right) \approx \frac{1}{N_s}\sum_{m=1}^{N_s} N\left(\boldsymbol{\lambda}_r \mid \boldsymbol{\mu}_\lambda^{(m)} + \mathbf{K}_\lambda^{(m)}\left(\boldsymbol{\mu}_\lambda^{(m)} - \hat{\boldsymbol{\lambda}}_r\right), \boldsymbol{\Sigma}_{\lambda\lambda}^{(m)} - \mathbf{K}_\lambda^{(m)}\boldsymbol{\Sigma}_{\lambda\lambda}^{(m)}\right) \qquad (28)$$

where $\boldsymbol{\mu}_\lambda^{(m)}$ and $\boldsymbol{\Sigma}_{\lambda\lambda}^{(m)}$ denote the samples taken from $p(\boldsymbol{\mu}_\lambda, \boldsymbol{\Sigma}_{\lambda\lambda} | \{D_s\}_{s=1,s\neq r}^{N_D})$, and $\mathbf{K}_\lambda^{(m)} = \boldsymbol{\Sigma}_{\lambda\lambda}^{(m)} (\boldsymbol{\Sigma}_{\lambda\lambda}^{(m)} + \hat{\boldsymbol{\Sigma}}_{\lambda_r \lambda_r})^{-1}$ is the sampled gain matrix. It can be proved that for this Gaussian mixture distribution the mean and covariance matrix can be computed as [61]

$$E(\boldsymbol{\lambda}_r) = \frac{1}{N_s} \sum_{m=1}^{N_s} \left( \boldsymbol{\mu}_\lambda^{(m)} + \mathbf{K}_\lambda^{(m)} (\boldsymbol{\mu}_\lambda^{(m)} - \hat{\boldsymbol{\lambda}}_r) \right) \tag{29}$$

$$C(\boldsymbol{\lambda}_r) = \frac{1}{N_s} \sum_{m=1}^{N_s} \left( \left( \boldsymbol{\mu}_\lambda^{(m)} + \mathbf{K}_\lambda^{(m)} (\boldsymbol{\mu}_\lambda^{(m)} - \hat{\boldsymbol{\lambda}}_r) \right) \left( \boldsymbol{\mu}_\lambda^{(m)} + \mathbf{K}_\lambda^{(m)} (\boldsymbol{\mu}_\lambda^{(m)} - \hat{\boldsymbol{\lambda}}_r) \right)^T + \boldsymbol{\Sigma}_{\lambda\lambda}^{(m)} - \mathbf{K}_\lambda^{(m)} \boldsymbol{\Sigma}_{\lambda\lambda}^{(m)} \right)$$
$$- \left( \frac{1}{N_s} \sum_{m=1}^{N_s} \left( \boldsymbol{\mu}_\lambda^{(m)} + \mathbf{K}_\lambda^{(m)} (\boldsymbol{\mu}_\lambda^{(m)} - \hat{\boldsymbol{\lambda}}_r) \right) \right) \left( \frac{1}{N_s} \sum_{m=1}^{N_s} \left( \boldsymbol{\mu}_\lambda^{(m)} + \mathbf{K}_\lambda^{(m)} (\boldsymbol{\mu}_\lambda^{(m)} - \hat{\boldsymbol{\lambda}}_r) \right) \right)^T \tag{30}$$

where $E(\boldsymbol{\lambda}_r)$ and $C(\boldsymbol{\lambda}_r)$ denote the mean vector and covariance matrix, respectively. Likewise, the integral in Eq. (26) can be calculated as

$$p(\boldsymbol{\lambda}_{N_D+1} | \{D_s\}_{s=1}^{N_D}) \approx \frac{1}{N_s} \sum_{l=1}^{N_s} N(\boldsymbol{\lambda}_{N_D+1} | \boldsymbol{\mu}_\lambda^{(l)}, \boldsymbol{\Sigma}_{\lambda\lambda}^{(l)}) \tag{31}$$

where $\boldsymbol{\mu}_\lambda^{(l)}$ and $\boldsymbol{\Sigma}_{\lambda\lambda}^{(l)}$ are the samples taken from $p(\boldsymbol{\mu}_\lambda, \boldsymbol{\Sigma}_{\lambda\lambda} | \{D_s\}_{s=1}^{N_D})$. Given this approximation, the mean and covariance of $\boldsymbol{\lambda}_{N_D+1}$ can be calculated similar to Eqs. (29-30) giving [61]:

$$E(\boldsymbol{\lambda}_{N_D+1}) = \frac{1}{N_s} \sum_{l=1}^{N_s} \boldsymbol{\mu}_\lambda^{(l)} \tag{32}$$

$$C(\boldsymbol{\lambda}_{N_D+1}) = \frac{1}{N_s} \sum_{l=1}^{N_s} \left( \boldsymbol{\mu}_\lambda^{(l)} \boldsymbol{\mu}_\lambda^{(l)T} + \boldsymbol{\Sigma}_\lambda^{(l)} \right) - \left( \frac{1}{N_s} \sum_{l=1}^{N_s} \boldsymbol{\mu}_\lambda^{(l)} \right) \left( \frac{1}{N_s} \sum_{l=1}^{N_s} \boldsymbol{\mu}_\lambda^{(l)} \right)^T \tag{33}$$

where $E(\boldsymbol{\lambda}_{N_D+1})$ and $C(\boldsymbol{\lambda}_{N_D+1})$ are the mean vector and covariance matrix of $\boldsymbol{\lambda}_{N_D+1}$, respectively.

The formulations presented in this section are systematized in Algorithm 1. This algorithm is called combined asymptotic approximation and sampling method since it employs the asymptotic approximation to identify the dynamical parameters corresponding to each experiment and then establishes an MCMC sampling scheme for updating of the hyper-parameters. The proposed MCMC approach leads to reasonable approximations independent of the physical model runs. However, the computational cost increases as the number of dynamical parameters and data sets grow.

**Algorithm 1**
Combined asymptotic approximation and MCMC sampling method

1. Carry out the Bayesian FFT modal identification
   For each data set $D_s$, where $s = \{1, 2, ..., N_D\}$, do the following:
   1.1. Minimize $L(\boldsymbol{\theta}_s)$ given by Eq. (9) with respect to $\boldsymbol{\theta}_s$
   1.2. Compute the MPV ($\hat{\boldsymbol{\theta}}_s$) using Eq. (10)
   1.3. Compute the covariance matrix $\hat{\boldsymbol{\Sigma}}_{\boldsymbol{\theta}_s \boldsymbol{\theta}_s}$ using Eq. (11)
   1.4. Select the dynamical parameters and their associated uncertainty $\hat{\boldsymbol{\lambda}}_s$'s and $\hat{\boldsymbol{\Sigma}}_{\boldsymbol{\lambda}_s \boldsymbol{\lambda}_s}$'s
   End for
2. Quantify the uncertainty associated with the dynamical parameters of $r^{\text{th}}$ test
   For each data set $D_r$, where $r = \{1, 2, ..., N_D\}$, do the following:
   2.1. Compute the marginal distribution $p\left(\boldsymbol{\mu}_\lambda, \boldsymbol{\Sigma}_{\lambda\lambda} \mid \{D_s\}_{s=1, s \neq r}^{N_D}\right)$ using Eq. (22)
   2.2. Draw samples $\boldsymbol{\mu}_\lambda^{(m)}$ and $\boldsymbol{\Sigma}_{\lambda\lambda}^{(m)}$, $m = \{1, 2, ..., N_s\}$, from $p\left(\boldsymbol{\mu}_\lambda, \boldsymbol{\Sigma}_{\lambda\lambda} \mid \{D_s\}_{s=1, s \neq r}^{N_D}\right)$
   2.3. Compute the posterior distribution $p\left(\boldsymbol{\lambda}_r \mid \{D_s\}_{s=1}^{N_D}\right)$ using Eq. (28)
   2.4. Estimate the mean vector of $\boldsymbol{\lambda}_r$ using Eq. (29)
   2.5. Estimate the covariance matrix of $\boldsymbol{\lambda}_r$ using Eq. (30)
   End for
3. Propagate the uncertainty for unobserved system operating conditions
   3.1. Compute the hyper-parameters marginal distribution $p\left(\boldsymbol{\mu}_\lambda, \boldsymbol{\Sigma}_{\lambda\lambda} \mid \{D_s\}_{s=1}^{N_D}\right)$ by Eq. (20)
   3.2. Draw the samples $\boldsymbol{\mu}_\lambda^{(m)}$ and $\boldsymbol{\Sigma}_{\lambda\lambda}^{(m)}$, $m = \{1, 2, ..., N_s\}$, from $p\left(\boldsymbol{\mu}_\lambda, \boldsymbol{\Sigma}_{\lambda\lambda} \mid \{D_s\}_{s=1}^{N_D}\right)$
   3.3. Compute the posterior distribution $p\left(\boldsymbol{\lambda}_{N_D+1} \mid \{D_s\}_{s=1}^{N_D}\right)$ using Eq. (31)
   3.4. Estimate the mean vector of $\boldsymbol{\lambda}_{N_D+1}$ using Eq. (32)
   3.5. Estimate the covariance matrix of $\boldsymbol{\lambda}_{N_D+1}$ using Eq. (33)

### 4.2. Dual Laplace approximation approach

*4.2.1. MAP estimations of the hyper-parameters*

In this section, an efficient method is proposed to compute the hyper-parameters MAP estimations. This necessitates maximizing the hyper-parameters marginal distribution given in Eq. (20). However, it is preferred to minimize the negative logarithm of the marginal distribution given by

$$L(\boldsymbol{\mu}_\lambda, \boldsymbol{\Sigma}_{\lambda\lambda}) \triangleq -\ln p\left(\boldsymbol{\mu}_\lambda, \boldsymbol{\Sigma}_{\lambda\lambda} \mid \{D_s\}_{s=1}^{N_D}\right)$$
$$= \frac{1}{2} \sum_{s=1}^{N_D} \ln \left| \det\left(\boldsymbol{\Sigma}_{\lambda\lambda} + \hat{\boldsymbol{\Sigma}}_{\lambda_s \lambda_s}\right)\right| + \frac{1}{2} \sum_{s=1}^{N_D} \left[(\boldsymbol{\mu}_\lambda - \hat{\boldsymbol{\lambda}}_s)^T (\boldsymbol{\Sigma}_{\lambda\lambda} + \hat{\boldsymbol{\Sigma}}_{\lambda_s \lambda_s})^{-1} (\boldsymbol{\mu}_\lambda - \hat{\boldsymbol{\lambda}}_s)\right] + c' \quad (34)$$

For a reliable optimization, it is ideal to feed analytical derivatives into the optimization toolbox. Analytical derivatives of this objective function are obtained in Appendix (C), which gives:

$$\frac{\partial L(\boldsymbol{\mu}_\lambda, \boldsymbol{\Sigma}_{\lambda\lambda})}{\partial \boldsymbol{\mu}_\lambda} = \sum_{s=1}^{N_D}\left[(\boldsymbol{\Sigma}_{\lambda\lambda} + \hat{\boldsymbol{\Sigma}}_{\lambda_s\lambda_s})^{-1}(\boldsymbol{\mu}_\lambda - \hat{\boldsymbol{\lambda}}_s)\right] \tag{35}$$

$$\frac{\partial L(\boldsymbol{\mu}_\lambda, \boldsymbol{\Sigma}_{\lambda\lambda})}{\partial \boldsymbol{\Sigma}_{\lambda\lambda}} = \frac{1}{2}\sum_{s=1}^{N_D}\left[\left(\boldsymbol{\Sigma}_{\lambda\lambda} + \hat{\boldsymbol{\Sigma}}_{\lambda_s\lambda_s}\right)^{-1} - \left(\boldsymbol{\Sigma}_{\lambda\lambda} + \hat{\boldsymbol{\Sigma}}_{\lambda_s\lambda_s}\right)^{-1}\left(\boldsymbol{\mu}_\lambda - \hat{\boldsymbol{\lambda}}_s\right)\left(\boldsymbol{\mu}_\lambda - \hat{\boldsymbol{\lambda}}_s\right)^T \left(\boldsymbol{\Sigma}_{\lambda\lambda} + \hat{\boldsymbol{\Sigma}}_{\lambda_s\lambda_s}\right)^{-1}\right] \tag{36}$$

When the gradient vector in Eq. (35) is set to zero, one obtains:

$$\hat{\boldsymbol{\mu}}_\lambda = \sum_{s=1}^{N_D}\boldsymbol{\Lambda}_s\hat{\boldsymbol{\lambda}}_s \ ; \quad \boldsymbol{\Lambda}_s = \left[\sum_{r=1}^{N_D}\left(\boldsymbol{\Sigma}_{\lambda\lambda} + \hat{\boldsymbol{\Sigma}}_{\lambda_r\lambda_r}\right)^{-1}\right]^{-1}\left(\boldsymbol{\Sigma}_{\lambda\lambda} + \hat{\boldsymbol{\Sigma}}_{\lambda_s\lambda_s}\right)^{-1} \tag{37}$$

where $\hat{\boldsymbol{\mu}}_\lambda$ is the MAP estimation of $\boldsymbol{\mu}_\lambda$ and $\boldsymbol{\Lambda}_s$ is the weighting matrix. This explicit solution aids to express all unknown parameters in terms of the hyper covariance matrix. It also suggests that those data sets containing a smaller number of data points (potentially larger uncertainty) will be given smaller weights since $\boldsymbol{\Lambda}_s$ is inversely proportional to $\hat{\boldsymbol{\Sigma}}_{\lambda_s\lambda_s}$.

To carry out the minimization, initial estimations of the hyper-parameters are required. To obtain an initial estimation, one may consider equal covariance matrices for all data sets, i.e. $\hat{\boldsymbol{\Sigma}}_{\lambda_s\lambda_s} = \hat{\boldsymbol{\Sigma}}_0, \forall s = 1, 2, ..., N_D$. This assumption holds when the model perfectly describes the actual response, which leads to the following initial estimations:

$$\bar{\boldsymbol{\mu}}_\lambda = \frac{1}{N_D}\sum_{s=1}^{N_D}\hat{\boldsymbol{\lambda}}_s \tag{38}$$

$$\bar{\boldsymbol{\Sigma}}_{\lambda\lambda} = \frac{1}{N_D}\sum_{s=1}^{N_D}(\bar{\boldsymbol{\mu}}_\lambda - \hat{\boldsymbol{\lambda}}_s)(\bar{\boldsymbol{\mu}}_\lambda - \hat{\boldsymbol{\lambda}}_s)^T - \hat{\boldsymbol{\Sigma}}_0 \tag{39}$$

where $\bar{\boldsymbol{\mu}}_\lambda$ and $\bar{\boldsymbol{\Sigma}}_{\lambda\lambda}$ are approximate mean and covariance, respectively. Note that a reasonable choice for $\hat{\boldsymbol{\Sigma}}_0$ is the mean of covariance matrices expressed as $\sum_{s=1}^{N_D}\hat{\boldsymbol{\Sigma}}_{\lambda_s\lambda_s}/N_D$.

Aside from the primary use of these estimations, they provide conceptual and practical insights on the outcome of the proposed hierarchical framework. For example, when the covariance matrix $\hat{\boldsymbol{\Sigma}}_0$ is disregarded from Eq. (39), these estimations exactly coincide with the ensemble mean and covariance matrix computed while following a Frequentist approach. On the contrary, these results are attained following a holistic Bayesian approach. When $\hat{\boldsymbol{\Sigma}}_0$ is included in Eq. (39), the covariance matrix estimated is essentially smaller than the ensemble covariance matrix. This reduction of the uncertainty suggests that the proposed framework attributes smaller portion to the variability over data sets as compared to the Frequentist approach. The importance of these findings is in removing any need to incorporate Frequentist concepts for considering the variability.

Once the hyper-parameters MAP estimation is computed, the integrals appeared in Eqs. (25) and (26) can be approximated as

$$p\left(\boldsymbol{\lambda}_r \mid \{D_s\}_{s=1}^{N_D}\right) \approx N\left(\boldsymbol{\lambda}_r \mid \hat{\boldsymbol{\lambda}}_r + \hat{\mathbf{K}}_{\lambda_r}\left(\hat{\boldsymbol{\mu}}_\lambda - \hat{\boldsymbol{\lambda}}_r\right), \hat{\boldsymbol{\Sigma}}_{\lambda_r \lambda_r} - \hat{\mathbf{K}}_{\lambda_r} \hat{\boldsymbol{\Sigma}}_{\lambda_r \lambda_r}\right) \qquad (40)$$

$$p\left(\boldsymbol{\lambda}_{N_D+1} \mid \{D_s\}_{s=1}^{N_D}\right) \approx N\left(\boldsymbol{\lambda}_{N_D+1} \mid \hat{\boldsymbol{\mu}}_\lambda, \hat{\boldsymbol{\Sigma}}_{\lambda\lambda}\right) \qquad (41)$$

where $\hat{\boldsymbol{\mu}}_\lambda$ and $\hat{\boldsymbol{\Sigma}}_{\lambda\lambda}$ are the MAP estimations, and $\hat{\mathbf{K}}_{\lambda_r} = \hat{\boldsymbol{\Sigma}}_{\lambda_r \lambda_r}\left(\hat{\boldsymbol{\Sigma}}_{\lambda_r \lambda_r} + \hat{\boldsymbol{\Sigma}}_{\lambda\lambda}\right)^{-1}$ is the optimal gain matrix. Notice that in derivation of these equations the uncertainty of the hyper-parameters is neglected and the MAP estimations are replaced into the integrand formulations. Moreover, the MAP estimation of the $p\left(\boldsymbol{\mu}_\lambda, \boldsymbol{\Sigma}_{\lambda\lambda} \mid \{D_s\}_{s=1}^{N_D}\right)$ is considered to be the same as $p\left(\boldsymbol{\mu}_\lambda, \boldsymbol{\Sigma}_{\lambda\lambda} \mid \{D_s\}_{s=1, s \neq r}^{N_D}\right)$. These conditions are often satisfied when there is a large number of data set.

The latest two equations are important since they share two different interpretations about the parametric uncertainty. Eq. (40) combines the information from multiple data sets so as to quantify the uncertainty involved with the dynamical parameters representing data set $D_r$. It modifies the estimated mean from $\hat{\boldsymbol{\lambda}}_r$ to $\hat{\boldsymbol{\lambda}}_r + \hat{\mathbf{K}}_\lambda\left(\hat{\boldsymbol{\mu}}_\lambda - \hat{\boldsymbol{\lambda}}_r\right)$ and explains it in terms of the deviation of $\hat{\boldsymbol{\lambda}}_r$ from $\hat{\boldsymbol{\mu}}_\lambda$ weighted by the gain matrix $\hat{\mathbf{K}}_\lambda$. Moreover, in Eq. (40), since the covariance matrix is expressed as the subtract of $\hat{\boldsymbol{\Sigma}}_{\lambda_r \lambda_r}$ and $\hat{\mathbf{K}}_\lambda \hat{\boldsymbol{\Sigma}}_{\lambda_r \lambda_r}$, the computed covariance matrix is essentially smaller than the covariance matrix computed using the data set $D_r$. This finding explains how using multiple data sets reduces the uncertainty associated with the dynamical parameters corresponding to a particular experiment (data set). In contrast, Eq. (41) gives the posterior predictive distribution of the dynamical parameters approximated by the MAP estimation of the hyper covariance matrix. While the former interpretation of uncertainty indicates the identification precision associated with the parameters of a particular experiment, the latter solely accounts for the variability of parameters enforced due to modeling errors since the uncertainty of the hyper-parameters are neglected. These results also indicate how the hierarchical approach separates different types of uncertainty and provides a broader standpoint for the uncertainty quantification in modal identification problems. A notable feature of the presented framework is that it avoids trapping in philosophical controversies existing between the Frequentist and Bayesian statisticians, and at the same time satisfies the Bayesian probability legitimates that includes the fundamental probability theory axioms.

*4.2.2. Uncertainty of the hyper-parameters*

When the hyper-parameters are globally-identifiable, for a large number of data sets the marginal posterior distribution of the hyper-parameters can be well approximated using the Laplace asymptotic approximation [14,58]. Using Taylor series expansion around the MAP estimation provides the following second-order approximation for the LHS of Eq. (34) giving:

$$L(\mathbf{\psi}) \approx L(\hat{\mathbf{\psi}}) + \frac{1}{2}(\mathbf{\psi} - \hat{\mathbf{\psi}})^T \mathbf{H}(\mathbf{\psi})\big|_{\mathbf{\psi}=\hat{\mathbf{\psi}}} (\mathbf{\psi} - \hat{\mathbf{\psi}}) \qquad (42)$$

where $\mathbf{\psi} \in \mathbb{R}^{n(n+3)/2}$ denotes a vector comprising all unknown parameters of the hyper mean and covariance ($\mathbf{\mu}_\lambda$ and $\mathbf{\Sigma}_{\lambda\lambda}$); $\hat{\mathbf{\psi}} \in \mathbb{R}^{n(n+3)/2}$ denotes the MAP estimation of $\mathbf{\psi}$ computed in the preceding section; $L(\mathbf{\psi})$ denotes the second-order approximation of $L(\mathbf{\mu}_\lambda, \mathbf{\Sigma}_{\lambda\lambda})$; $\mathbf{H}(\hat{\mathbf{\psi}}) \in \mathbb{R}^{[n(n+3)/2] \times [n(n+3)/2]}$ is the Hessian matrix of $L(\mathbf{\mu}_\lambda, \mathbf{\Sigma}_{\lambda\lambda})$ calculated with respect to the elements of the hyper-parameters. Note that for determining the dimension of $\mathbf{\psi}$ the symmetry of the hyper covariance matrix ($\mathbf{\Sigma}_{\lambda\lambda}$) is postulated. For computing the Hessian matrix in an explicit manner, the reader is referred to Appendix (C). The inverse of this Hessian matrix evaluated at the MAP estimation ($\hat{\psi}$) yields the approximate covariance matrix. Ultimately, the following approximate posterior distribution can be attained:

$$p(\psi \mid \mathbf{D}) \approx N\left( \psi \mid \hat{\psi} , \left[ \mathbf{H}(\psi)\big|_{\psi=\hat{\psi}} \right]^{-1} \right) \tag{43}$$

where $p(\psi \mid \mathbf{D})$ is the Gaussian approximation for the hyper-parameters. Although this approximation might not hold in general, it can provide efficient approximations for the hyper-parameters uncertainty to be used as a means to check the validity of approximation presented in Eqs. (40-41).

Algorithm 2 outlines the formulations presented under the latest two sections. This algorithm is referred to as the dual asymptotic approximation method as it employs the Laplace asymptotic approximation for the uncertainty quantification of both data-set-specific parameters and the hyper-parameters.

**Algorithm 2**
Dual Laplace approximation method

1. Carry out the Bayesian FFT-based modal identification
   For each data set $D_s$, where $s = \{1, 2, ..., N_D\}$, do the following:
   1.1. Minimize $L(\boldsymbol{\theta}_s)$ given by Eq. (9) with respect to both $\boldsymbol{\theta}_s$
   1.2. Compute the MPV ($\hat{\boldsymbol{\theta}}_s$) using Eq. (10)
   1.3. Compute the covariance matrix $\hat{\boldsymbol{\Sigma}}_{\theta_s \theta_s}$ using Eq. (11)
   1.4. Select the dynamical parameters and their associated uncertainty ($\hat{\boldsymbol{\lambda}}_s$'s and $\hat{\boldsymbol{\Sigma}}_{\lambda_s \lambda_s}$'s) from the $\hat{\boldsymbol{\theta}}_s$ and $\hat{\boldsymbol{\Sigma}}_{\theta_s \theta_s}$
   End for
2. Compute the MAP estimations of the hyper-parameters
   2.1. Compute the hyper-parameters marginal distribution $p\left(\boldsymbol{\mu}_\lambda, \boldsymbol{\Sigma}_{\lambda\lambda} \mid \{D_s\}_{s=1}^{N_D}\right)$ by Eq. (20)
   2.2. Compute the initial estimations of the hyper-parameters ($\bar{\boldsymbol{\mu}}_\lambda$ and $\bar{\boldsymbol{\Sigma}}_{\lambda\lambda}$) using Eqs. (38-39) by setting $\hat{\boldsymbol{\Sigma}}_0 = \sum_{s=1}^{N_D} \hat{\boldsymbol{\Sigma}}_{\lambda_s \lambda_s} / N_D$
   2.3. Minimize $L(\boldsymbol{\mu}_\lambda, \boldsymbol{\Sigma}_{\lambda\lambda})$ using the analytical gradient vector given using Eqs. (35-36) and compute the MAP estimations $\hat{\boldsymbol{\mu}}_\lambda$ and $\hat{\boldsymbol{\Sigma}}_{\lambda\lambda}$
   2.4. Compute the Hessian inverse of $L(\boldsymbol{\mu}_\lambda, \boldsymbol{\Sigma}_{\lambda\lambda})$ evaluated at the MAP estimations
3. Quantify the uncertainty associated with the dynamical parameters of $r^{th}$ test
   For each data set $D_r$, where $r = \{1, 2, ..., N_D\}$, do the following:
   3.1. Compute the posterior distribution $p\left(\boldsymbol{\lambda}_r \mid \{D_s\}_{s=1}^{N_D}\right)$ using Eq. (40)
   End for
4. Propagate the uncertainty for an unobserved system operating condition
   4.1. Compute the posterior distribution $p\left(\boldsymbol{\lambda}_{N_D+1} \mid \{D_s\}_{s=1}^{N_D}\right)$ using Eq. (41)

### 4.3. Identifiability of the hyper covariance matrix

The procedure suggested earlier for sampling either diagonal or non-diagonal covariance matrices reduces the number of involved parameters and delivers positive semi-definite covariance matrices. Nevertheless, in the case that the off-diagonal elements are unidentifiable, the statistical moments can vary drastically over each trial of the MCMC samplers. This unidentifiability prevails while using the Laplace approximation method since the solution shows a high sensitivity to the initial estimations. This identifiability problem also involves the samples of the diagonal elements to some extent. Due to such problems, recent studies on hierarchical Bayesian approach (e.g. [44,47]) have considered the hyper covariance matrix to be diagonal and neglected the correlation between the parameters. This approach is inapplicable for modal identification problems as the components of mode shape vectors could be highly correlated.

In this study, we suggest a simple remedy using the eigenspace representation of the covariance matrix and seek to resolve this problem through separating the identification of the eigenvectors from the eigenvalues. Thus, the covariance matrix is decomposed as $\mathbf{\Sigma}_{\lambda\lambda} = \mathbf{L}_{\lambda\lambda}\mathbf{D}_{\lambda\lambda}\mathbf{L}_{\lambda\lambda}^T$, where $\mathbf{L}_{\lambda\lambda} \in \mathbb{R}^{(n+2)\times(n+2)}$ is a unitary matrix whose columns contain the orthonormal eigenvectors and $\mathbf{D}_{\lambda\lambda} \in \mathbb{R}^{(n+2)\times(n+2)}$ is a diagonal matrix containing the eigenvalues corresponding to each column of $\mathbf{L}_{\lambda\lambda}$. The idea is to consider reasonable estimations for the eigenvectors ($\mathbf{L}_{\lambda\lambda}$), while the eigenvalues ($\mathbf{D}_{\lambda\lambda}$) are to be computed using either the sampling or the Laplace approximation method. To specify $\mathbf{L}_{\lambda\lambda}$, the approximate MAP estimation of the covariance matrix ($\overline{\mathbf{\Sigma}}_{\lambda\lambda}$) obtained in Eq. (39) is decomposed into its eigenspace representation $\overline{\mathbf{\Sigma}}_{\lambda\lambda} = \overline{\mathbf{L}}_{\lambda\lambda}\overline{\mathbf{D}}_{\lambda\lambda}\overline{\mathbf{L}}_{\lambda\lambda}^T$, and $\mathbf{L}_{\lambda\lambda}$ is assumed to be equal to $\overline{\mathbf{L}}_{\lambda\lambda}$. Therefore, the hyper covariance matrix is rewritten as $\mathbf{\Sigma}_{\lambda\lambda} = \overline{\mathbf{L}}_{\lambda\lambda}\mathbf{D}_{\lambda\lambda}\overline{\mathbf{L}}_{\lambda\lambda}^T$, in which the diagonal matrix ($\mathbf{D}_{\lambda\lambda}$) is to be estimated. Using this simplification assist to overcome the unidentifiability of the hyper-parameters even for small number of data sets since the number of unknown parameters of the hyper covariance matrix is reduced considerably. The flow of steps required for simplifying the hyper covariance matrix is summarized in Algorithm 3.

**Algorithm 3**
Simplification of the hyper covariance matrix using the eigenbasis decomposition

1. Compute the hyper-parameters marginal distribution $p\left(\mathbf{\mu}_\lambda, \mathbf{\Sigma}_{\lambda\lambda} \mid \{D_s\}_{s=1}^{N_D}\right)$ by Eq. (20)
2. Compute the initial estimations of the hyper-parameters ($\overline{\mathbf{\mu}}_\lambda$ and $\overline{\mathbf{\Sigma}}_{\lambda\lambda}$) using Eqs. (38-39) by setting $\hat{\mathbf{\Sigma}}_0 = \sum_{s=1}^{N_D} \hat{\mathbf{\Sigma}}_{\lambda_s\lambda_s} / N_D$
3. Compute the Cholesky decomposition as $\overline{\mathbf{\Sigma}}_{\lambda\lambda} = \overline{\mathbf{L}}_{\lambda\lambda}\overline{\mathbf{D}}_{\lambda\lambda}\overline{\mathbf{L}}_{\lambda\lambda}^T$
4. Reparametrize the hyper covariance matrix as $\mathbf{\Sigma}_{\lambda\lambda} = \overline{\mathbf{L}}_{\lambda\lambda}\mathbf{D}_{\lambda\lambda}\overline{\mathbf{L}}_{\lambda\lambda}^T$

**Remark-3.** The Laplace approximation simplifies the likelihood function and leads to an explicit formulation for $p\left(\boldsymbol{\mu}_\lambda, \boldsymbol{\Sigma}_{\lambda\lambda} | \{D_s\}_{s=1}^{N_D}\right)$. This formulation allows for considering more efficient and systematic options for updating the hyper-parameters. In this respect, the MCMC sampling over $p\left(\boldsymbol{\mu}_\lambda, \boldsymbol{\Sigma}_{\lambda\lambda} | \{D_s\}_{s=1}^{N_D}\right)$ is highly efficient since it is free of model evaluations. Moreover, the MAP estimations of $\boldsymbol{\mu}_\lambda$ and $\boldsymbol{\Sigma}_{\lambda\lambda}$ can also be computed without requiring model evaluations. On the contrary, full sampling strategies involve a large number of parameters (model parameters and prediction error parameters for each data set, as well as hyper-parameters) and require a large number of model evaluations, which dramatically increases the computational costs.

**Remark-4.** Extensive research efforts have been devoted to provide reliable algorithms for modal identification using the fast Bayesian FFT approach [20–22]. This includes the computation of the most probable values and the Hessian matrix under unit norm constraints. The Laplace approximation allows integrating these efficient algorithms within the hierarchical framework, while the full MCMC sampling will not offer such advantages.

**Remark-5.** We recall that the mode shape unit-norm constraints are considered in the Laplace approximation appearing in Eq. (12). Thus, when the samples of $\boldsymbol{\mu}_{\varphi_i}$ are drawn from either Eqs. (20) or (22), it turns out that they automatically satisfy the unit-norm constraints. The same holds when the MAP estimation of $\boldsymbol{\mu}_{\varphi_i}$ are computed. The reason for this issue should be searched in the particular form of the probability distribution appearing in Eqs. (20) and (22), where in the computation of the mean vectors ($\hat{\boldsymbol{\lambda}}_s$'s) and the covariance matrices ($\hat{\boldsymbol{\Sigma}}_{\lambda_s\lambda_s}$'s) the unit-norm conditions are implemented.

## 5. Illustrative examples

### 5.1. Three-story prototype structure

Fig. 2(a) shows the three-story shear building prototype structure tested on a shaking table. The dynamical properties of this prototype are reported in previous studies [60,63,64] that will be used as a benchmark for the validation of modal identification results. The acceleration time-history responses of the three stories were measured while the prototype was subjected to $N_D = 40$ independent GWN base excitations. Each set of the time-history acceleration measurements is 60s long sampled at 0.005s intervals. Note that the chosen length is long enough such that it allows using the Laplace approximation approach for inferring the parameters from each data set.

The averaged spectrum of the singular values is shown in Fig. 2(b). The three resonant peaks are well-separated, which potentially represent three dynamical modes. The frequency bands [3.2-5.2Hz], [12.0-14.0Hz], and [17.5-19.5Hz] are considered to infer the dynamical parameters using the frequency-domain responses. Algorithms 1 and 2 can be next employed to infer the dynamical parameters, to combine the information from multiple data sets, and to quantify the uncertainties. As the first step, the Laplace approximation is applied for identifying the parameters from each data set. This step is identical for both algorithms, which obtains the underlying parameters of the Gaussian approximations. Thus, the mean vectors ($\hat{\boldsymbol{\theta}}_s$'s) and the covariance matrices ($\hat{\boldsymbol{\Sigma}}_{\boldsymbol{\theta}_s \boldsymbol{\theta}_s}$'s) corresponding to the three dynamical modes are computed from each data set. Fig. 2(c) schematizes the three mode shapes identified from the data. The lower-diagonal plots in Figs. (3-5) visualizes the dynamical parameters ($\hat{\boldsymbol{\lambda}}_s$'s and $\hat{\boldsymbol{\Sigma}}_{\boldsymbol{\lambda}_s \boldsymbol{\lambda}_s}$'s) identified from different data sets. The blue circles indicate $\hat{\boldsymbol{\lambda}}_s$, and the associated error bars display one standard deviation margins. The large variability of these parameters is evident in this figure, which is to be addressed through the Algorithms 1 and 2. This task necessitates specifying the hyper-parameters prior distribution $p(\boldsymbol{\mu}_\lambda, \boldsymbol{\Sigma}_{\lambda\lambda})$. The elements of the hyper mean vector are described as

$$\mu_{f_i} \sim U(0, 25); \quad \mu_{\xi_i} \sim U(0, 0.1); \quad \mu_{\phi_{ij}} \sim U(-1, 1) \tag{44}$$

where $U(a,b)$ denotes the uniform probability distribution defined over the interval [a , b]. For the hyper covariance matrix, the eigenbasis transformation suggested earlier in Section 4.3 is used to reduce the number of involved parameters and to circumvent the potential unidentifiability of the hyper covariance matrix. Then, the uniform prior distribution $U(0, 0.1)$ is assigned to the eigenvalues of the hyper covariance matrix. To apply the MCMC sampling algorithm, the Transitional Markov chain Monte Carlo (TMCMC) sampling method [65,66] is applied for drawing the samples from the marginal posterior distribution of the hyper-parameters described by Eq. (20). The number of samples per each run of the TMCMC sampler is considered to be $N_s = 2000$ for all simulations. Given the TMCMC samples, the mean and covariance matrix of the parameters are then computed using Eqs. (32) and (33). Matrix plots of Figs. (3-5) depict the dynamical parameters and their associated uncertainties for the three dynamical modes. Those plots on the main diagonal show the Gaussian approximation of the marginal distribution of the parameters. The upper diagonal plots demonstrate the joint Gaussian approximation corresponding to each pair of the dynamical parameters. When these plots are compared with the lower-diagonal ones that show the variability of the parameters, it is revealed that the proposed hierarchical setting provides robust uncertainty with respect to multiple sources of uncertainty covering both the variability and identification precision. Introducing this new approach to uncertainty quantification is of great significance as it fills the existing gaps as to the quantification of the variability due to modeling errors commonly met in modal identification problems.

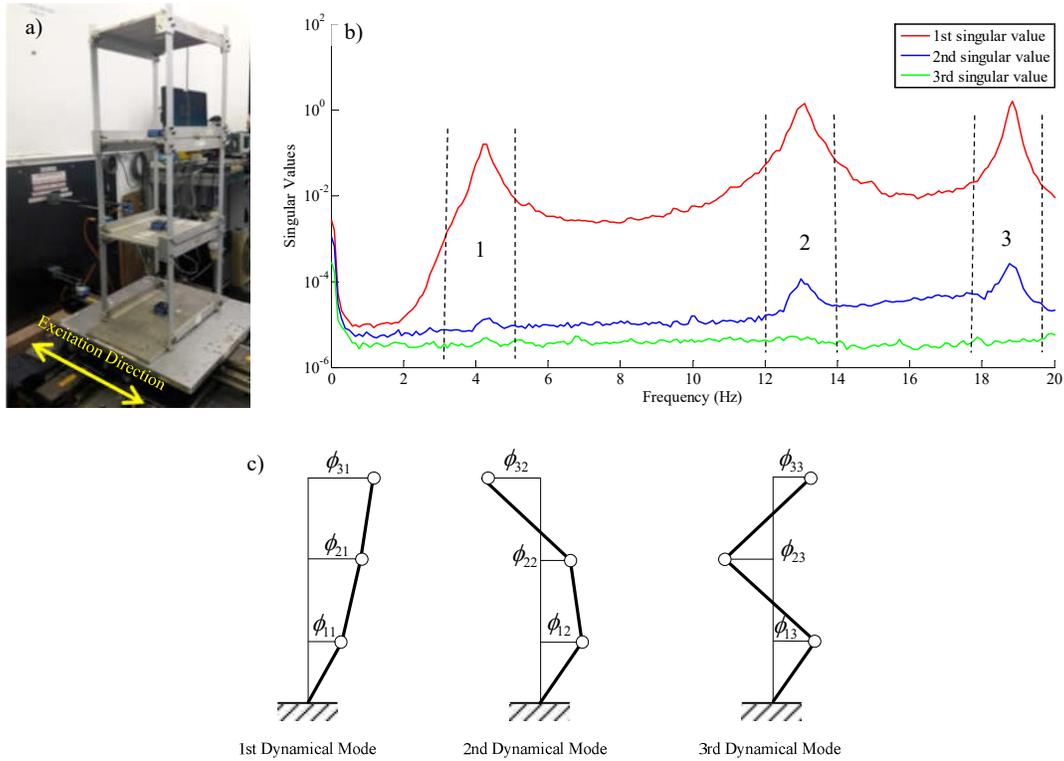

**Fig. 2.** (a) Prototype structure tested on a shaking table (b) Averaged singular value spectrum obtained from the three measured acceleration responses (c) Schematic overview of the dynamical modes

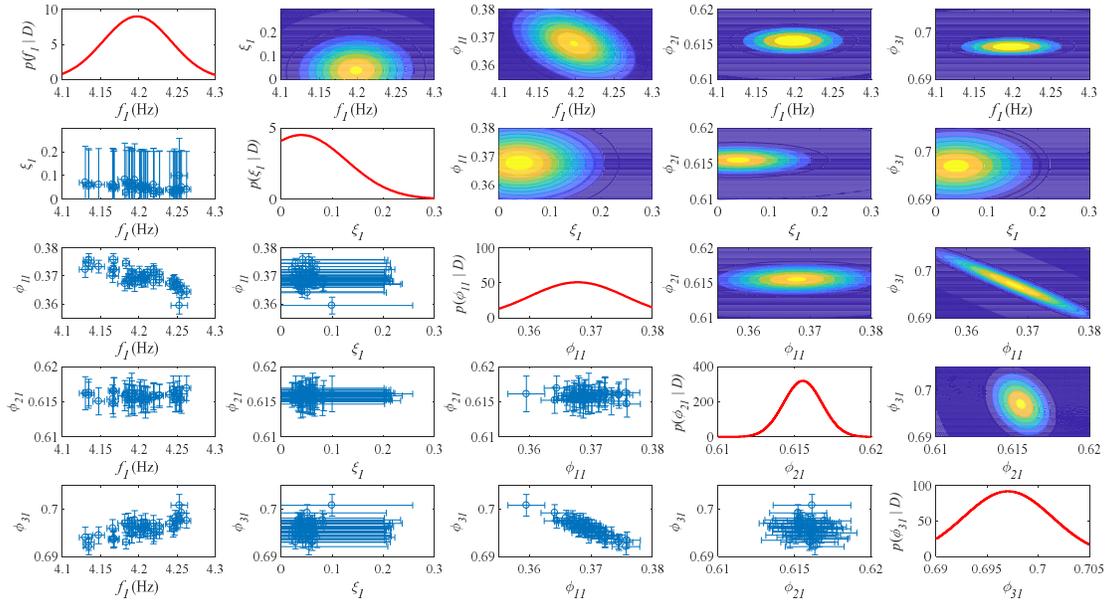

**Fig. 3.** Visualization of the uncertainty associated with the parameters of the 1$^{st}$ dynamical mode obtained using the MCMC sampling approach

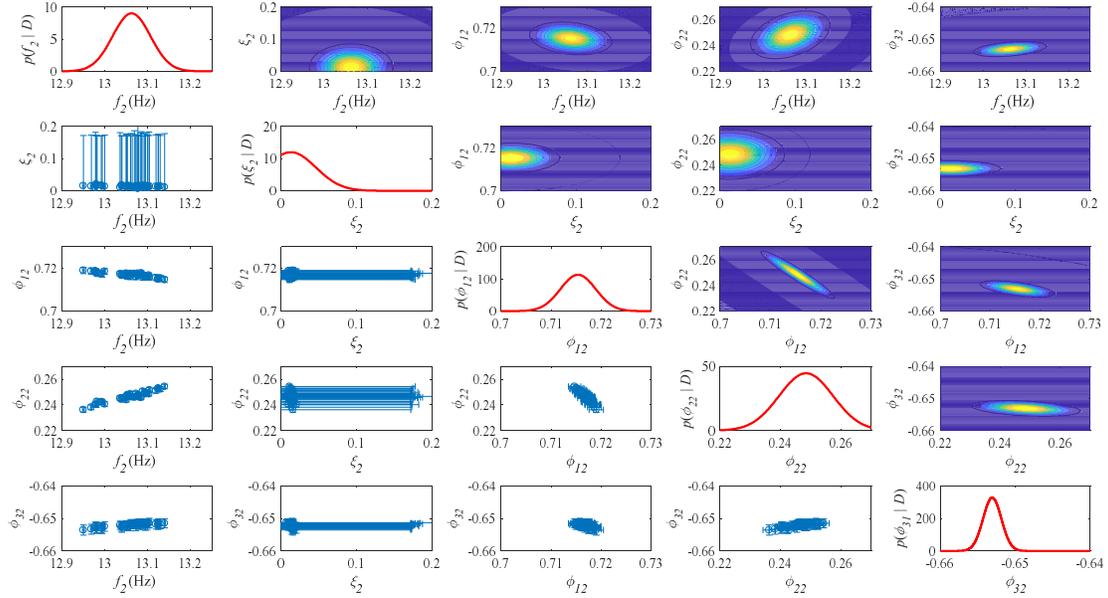

**Fig. 4.** Visualization of the uncertainty associated with the parameters of the 2$^{nd}$ dynamical mode obtained using the MCMC sampling approach

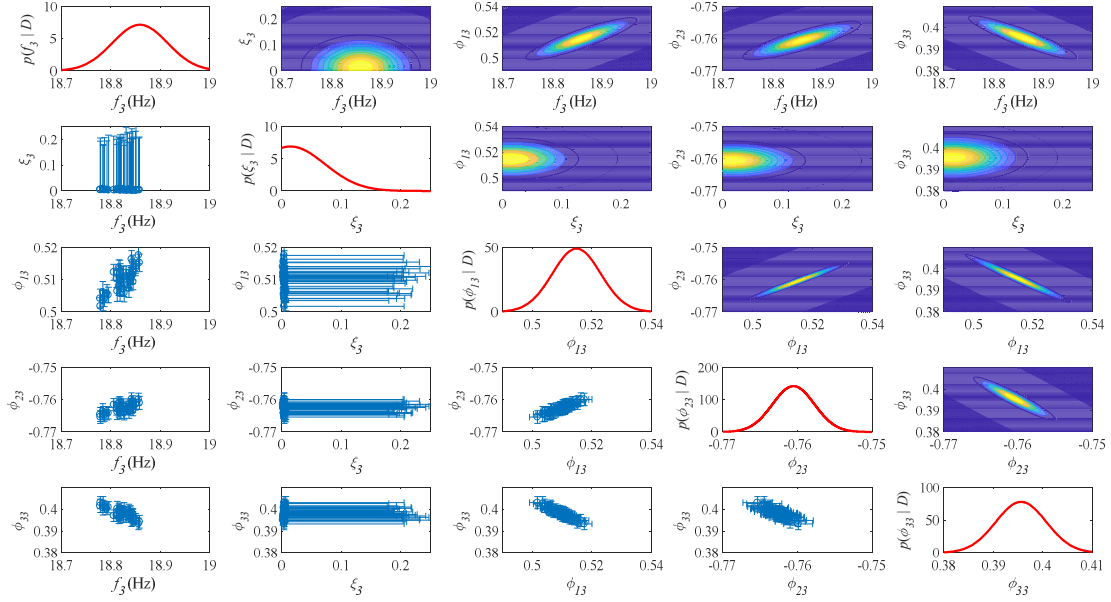

**Fig. 5.** Visualization of the uncertainty associated with the parameters of the 3$^{rd}$ dynamical mode obtained using the MCMC sampling approach

The Laplace approximation approach summarized in Algorithm 2 is a more efficient alternative to the sampling method. As the first step, the MAP estimations of the hyper-parameters are computed using the optimization process and approximations prescribed earlier. Table 1 compares the results of the two algorithms for the three dynamical modes. Results are in good agreement although the sampling method gives slightly larger uncertainty compared to the Laplace approximation approach. A noteworthy exception to this argument is the standard deviation of the modal damping ratios, which is assessed to be extremely large when the sampling approach is used. However, the Laplace approximation approach gives small standard deviations for the modal damping ratios. This issue is attributed to the fact that the MCMC sampling method captures the uncertainty of the hyper-parameters, whereas the Laplace asymptotic approximation approach neglects it completely. To predict such deviations, we may resort to the Hessian inverse of the hyper-parameters evaluated at the MAP estimations. In effect, we expect to encounter large uncertainty when the Laplace approximation is used for computing the uncertainty of the hyper-parameters. Table 2 presents the standard deviation attributed to the elements of the hyper mean vector and covariance matrix. As indicated, the standard deviation of all hyper-parameters is relatively small, whereas the standard deviation of $\mu_{\xi_i}$'s are extremely large. Thus, the simplified form of the Laplace approximation shall not be used when the hyper uncertainty is considerable, and this can simply be examined through the Hessian inverse. In effect, this notable difference is an indicator that assists to check whether the asymptotic approximation method remains valid.

It should be highlighted that the difference between the sampling and the asymptotic approximation methods might disappear as the number of observed data sets increases. In terms of the computational cost, the Laplace approximation method is greatly preferred, especially when the number of data sets is sufficiently large. For a small number of data sets (less than 10 data sets), the sampling approach is likely to be more accurate since it accounts for the uncertainty associated with the hyper-parameters.

**Table 1.**
Quantification of the uncertainty using the sampling and asymptotic approximation approaches

| Dynamical Parameters | Asymptotic Approximation | | Sampling Method | |
|---|---|---|---|---|
| | Mean | SD | Mean | SD |
| $f_1$ (Hz) | 4.205 | 0.0350 | 4.197 | 0.0444 |
| $\xi_1$ | 0.050 | 0.0014 | 0.039 | 0.0883 |
| $\phi_{11}$ | 0.370 | 0.0013 | 0.368 | 0.0033 |
| $\phi_{21}$ | 0.616 | 0.0022 | 0.615 | 0.0055 |
| $\phi_{31}$ | 0.696 | 0.0025 | 0.697 | 0.0063 |
| $f_2$ (Hz) | 13.078 | 0.0487 | 13.063 | 0.0442 |
| $\xi_2$ | 0.0135 | 0.0001 | 0.0130 | 0.0331 |
| $\phi_{12}$ | 0.717 | 0.0081 | 0.715 | 0.0069 |
| $\phi_{22}$ | 0.246 | 0.0028 | 0.248 | 0.0024 |
| $\phi_{32}$ | -0.651 | 0.0073 | -0.653 | 0.0063 |
| $f_3$ (Hz) | 18.824 | 0.0219 | 18.859 | 0.0557 |
| $\xi_3$ | 0.0047 | 0.0033 | 0.0147 | 0.0577 |
| $\phi_{13}$ | 0.510 | 0.0034 | 0.515 | 0.0051 |
| $\phi_{23}$ | -0.763 | 0.0051 | -0.761 | 0.0076 |
| $\phi_{33}$ | 0.398 | 0.0027 | 0.396 | 0.0040 |

**Table 2.**
Quantification of the uncertainty of the hyper-parameters using the second-order asymptotic approximation approach

| Hyper-parameters | Standard deviation of the hyper-parameters elements | | |
|---|---|---|---|
| | 1st Dynamical Mode ($i=1$) | 2nd Dynamical Mode ($i=2$) | 3rd Dynamical Mode ($i=3$) |
| $\mu_{f_i}$ | 0.0056 | 0.0080 | 0.0035 |
| $\mu_{\xi_i}$ | 0.0253 | 0.0256 | 0.0313 |
| $\mu_{\phi_{1i}}$ | 0.0006 | 0.0007 | 0.0008 |
| $\mu_{\phi_{2i}}$ | 0.0004 | 0.0016 | 0.0006 |
| $\mu_{\phi_{3i}}$ | 0.0004 | 0.0004 | 0.0007 |
| $\sigma_{f_i}$ | 0.0001 | 0.0003 | 5.4e-5 |
| $\sigma_{\xi_i}$ | 0.0040 | 0.0041 | 0.0061 |
| $\sigma_{\phi_{1i}}$ | 1.5e-6 | 3.1e-6 | 3.4e-6 |
| $\sigma_{\phi_{2i}}$ | 8.3e-7 | 1.6e-5 | 2.4e-6 |
| $\sigma_{\phi_{3i}}$ | 8.8e-7 | 1.0e-6 | 3.2e-6 |

## 5.2. Ambient vibration test

### 5.2.1. Description of the structure, sensing network, and measurements campaign

A single-tower cable footbridge located at the Hong Kong University of Science and Technology is selected to demonstrate the method. Fig. 6(a-b) shows longitudinal and elevation views of the bridge. The bridge is 22.5m long and 1.70m wide. Pedestrians' activities, wind actions, and the vehicles traveling from the underpass road can create significant vibrations. This enables us to have an easy-access structure for obtaining ambient vibration measurements. Triaxial Imote2$^©$ wireless accelerometers shown in Fig. 6(c) are used to measure the acceleration time-history responses. These sensors will shape a network composed of six leaf nodes, which communicates through a gateway node connected to the computer. The gateway node is shown in Fig. 6(d), and the network architecture is shown in Fig. 6(e). The synchronization of the sensors is carried out via the gateway node. To provide constant voltage over the experimentation, power banks are used to supply the electricity. Technical information about the calibration, setup, and implementation of the Imote2$^©$ sensors are obtained by the manufacturer in [67,68]. Fig. 6(f) shows the placement of the six sensors along the bridge. The sensors are distributed symmetrically across the bridge width. The shorter and longer spans are covered with 2 and 4 sensors, respectively. Nineteen sets of acceleration time-history responses ($N_D = 19$) were acquired. Each time-history data set is 5 min long sampled at 0.01s intervals. These measurements correspond to the ambient vibrations measured on a windy and cloudy day in March 2019 between 13:00-17:00 PM when the temperature was relatively constant around 21°C. Thus, the environmental conditions, such as temperature, are considered to be constant during the field experiments.

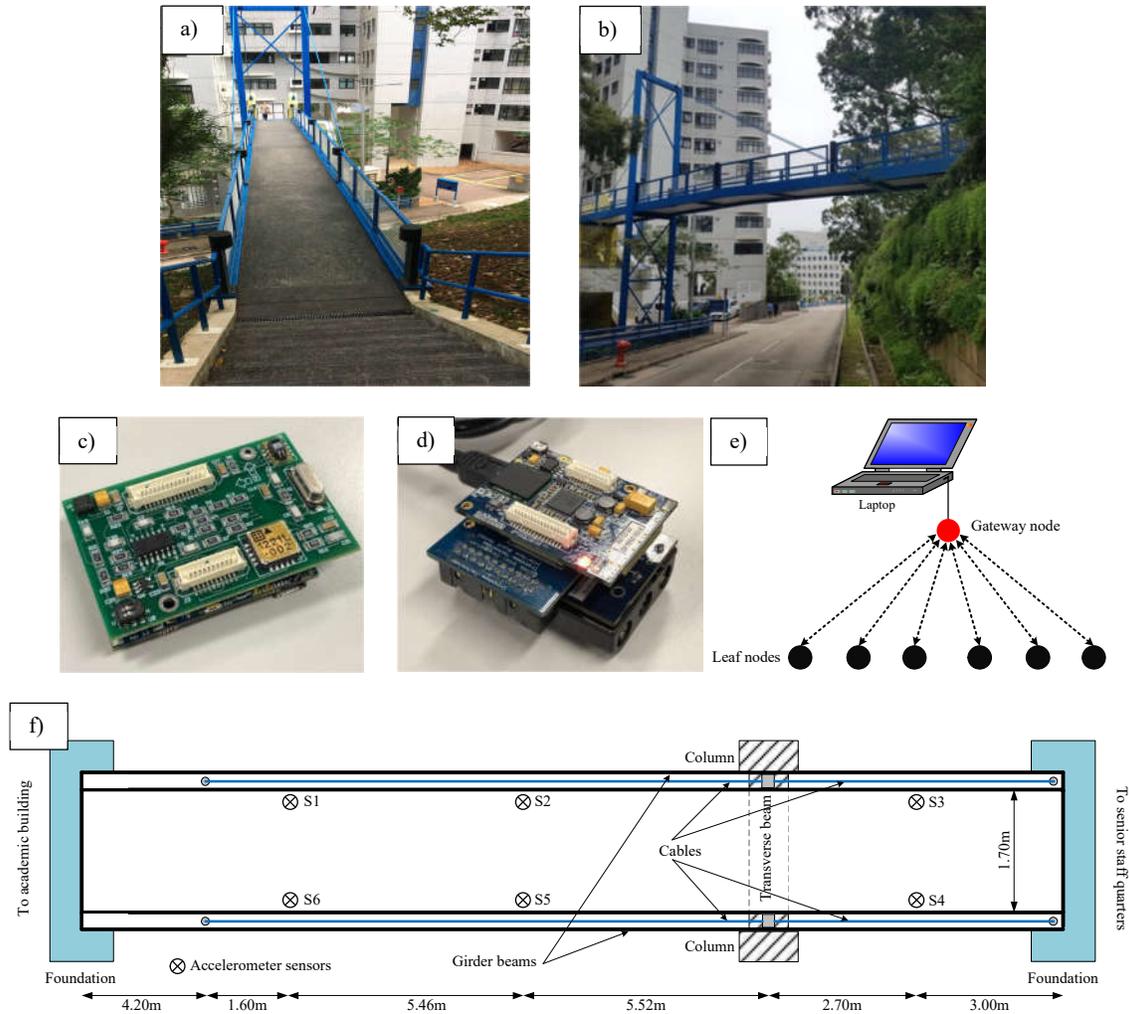

**Fig. 6.** (a-b) Longitudinal and elevation views the cable footbridge located at HKUST campus (c) Wireless sensor architecture (d) Imote2© wireless accelerometer acting as a leaf node (e) Getaway node used for communicating with the sensors (f) Floor plan of the bridge and the sensor placement configuration considered along the bridge

*5.2.2. Modal identification results*

Fig. 7 displays the averaged singular value spectrum obtained for the measured responses. The first four resonant peaks are indicated on this plot. The first four modal frequencies are observed to be at around 3.8Hz, 10.2Hz, 11.70Hz, and 20.0Hz, respectively. This result also agree with the modal identification exercised by Li and Chang [63]. Thus, the four non-overlapping frequency bands are considered to be [3.3-4.3 Hz], [9.7-10.7 Hz], [11.2-12.3 Hz], and [19.5-20.5Hz], respectively. Once the frequency bands are decided, we can perform the Bayesian modal identification to obtain the optimal estimations and the covariance matrices from each data set. Fig. 8(a-d) demonstrates the four identified modes. The longer span represents single and double curvatures in the first and second dynamical modes, respectively. The third mode shape is torsional, and the fourth mode is an unsymmetrical one. Although more dynamical modes could also be extracted from the data, in this study we only concentrate on these modes.

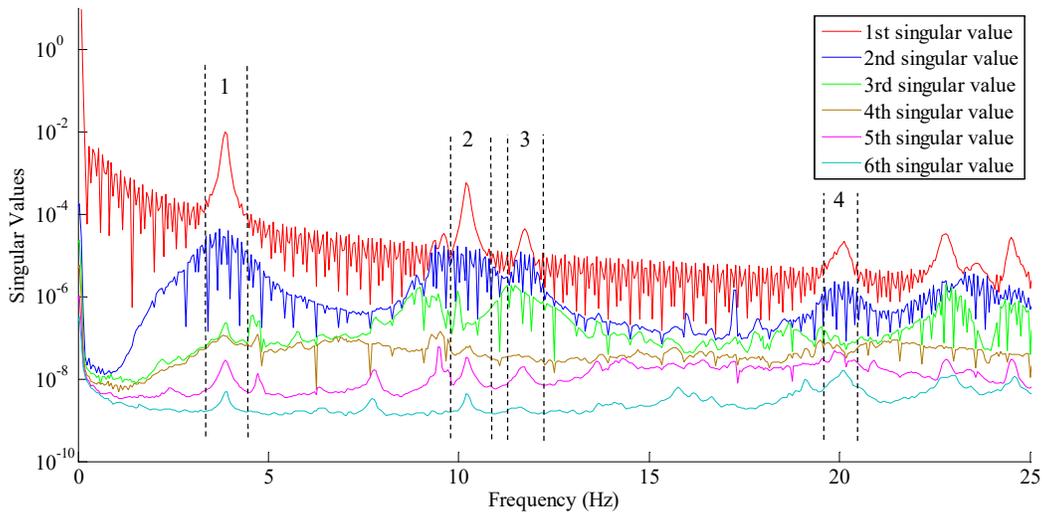

**Fig. 7.** Averaged singular value spectrum obtained from the acceleration response measurements (Dashed-lines indicate the selected resonant bands)

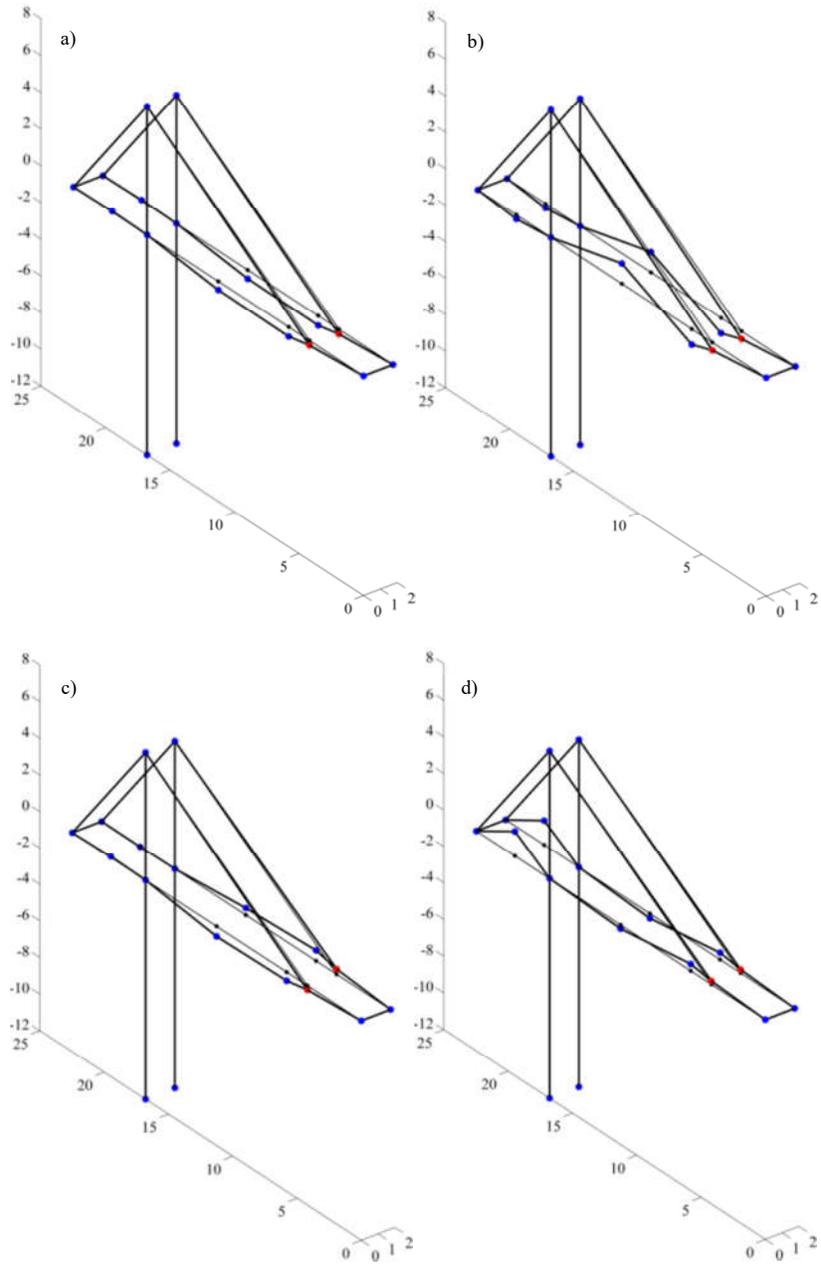

**Fig. 8.** Dynamical mode shapes identified using the Bayesian OMA approach corresponding to modal frequencies (a) $f_1 \approx$ 3.8Hz (b) $f_2 \approx 10.2$Hz (c) $f_3 \approx 11.7$Hz (d) $f_4 \approx 20.0$Hz

After computing the Gaussian approximation of the likelihood function corresponding to each data set, we marginalize the non-dynamical parameters as nuisance parameters and combine the dynamical parameters under the aforementioned hierarchical setting. As the number of data sets is relatively small, we tend to employ the sampling approach. The prior distribution of the hyper-parameters is described via uniform distributions chosen like the foregoing example.

Table 3 presents the mean and standard deviation of the dynamical parameters corresponding to the first four modes estimated using the sampling approach. It is demonstrated that the uncertainty of the dynamical parameters is considerable and can be attributed to both the identification precision and the variability. These uncertainties are far larger than the outcome of non-hierarchical Bayesian methods. Fig. 9(a) compares the marginal posterior distribution of the first modal frequency obtained using the hierarchical approach with the likelihood functions obtained from each data set. As shown, the realizations corresponding to each data set are sharply-peaked distributions, and the most probable values greatly vary from test-to-test, whereas the uncertainty bound associated with each data set is extremely small. Nevertheless, the hierarchical approach successfully accounts for this variability and yields robust uncertainty. For modal damping ratios, the opposite pattern can be observed in Fig. 9(b). The identification precision of the damping ratios obtained from each data set is relatively large owing to the errors existing in the assumptions about the damping characteristics. However, this type of uncertainty is reducible when multiple data sets are fused using the hierarchical approach. This reduction is reasonable and consistent with the fact that the identification uncertainty can be reduced as additional observations are employed. Note that results are qualitatively the same for the modal frequency and damping ratio, as well as the mode shape of the other three dynamical modes, but they are skipped here for brevity.

Table 3.
Probabilistic modal identification using the proposed sampling approach

| Dynamical Parameters | 1st Dynamical Mode | | 2nd Dynamical Mode | | 3rd Dynamical Mode | | 4th Dynamical Mode | |
|---|---|---|---|---|---|---|---|---|
| | Mean | SD | Mean | SD | Mean | SD | Mean | SD |
| $f_i$ (Hz) | 3.848 | 0.0462 | 10.206 | 0.0150 | 11.721 | 0.0483 | 20.055 | 0.0400 |
| $\xi_i$ | 0.014 | 0.0436 | 0.006 | 0.0793 | 0.018 | 0.0491 | 0.010 | 0.0177 |
| $\phi_{1i}$ | -0.514 | 0.0303 | -0.435 | 0.0263 | -0.531 | 0.0317 | -0.193 | 0.0128 |
| $\phi_{2i}$ | -0.467 | 0.0285 | 0.537 | 0.0325 | -0.485 | 0.0290 | 0.115 | 0.0076 |
| $\phi_{3i}$ | 0.049 | 0.0017 | -0.135 | 0.0082 | 0.034 | 0.0021 | -0.683 | 0.0453 |
| $\phi_{4i}$ | 0.049 | 0.0047 | -0.128 | 0.0078 | -0.039 | 0.0023 | -0.667 | 0.0443 |
| $\phi_{5i}$ | -0.477 | 0.0300 | 0.549 | 0.0332 | 0.424 | 0.0253 | 0.091 | 0.0061 |
| $\phi_{6i}$ | -0.526 | 0.0323 | -0.440 | 0.0266 | 0.549 | 0.0328 | -0.175 | 0.0116 |

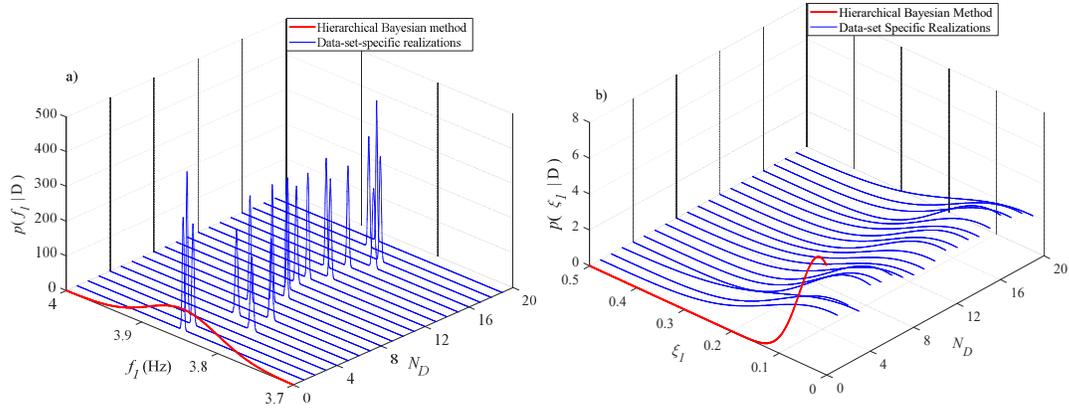

**Fig. 9.** (a-b) Comparison between the posterior predictive distribution and the data-set-specific likelihood functions for the modal frequency and damping of the first dynamical mode

It should be pointed out that the variability in the modal properties may exhibit specific correlation with the environmental factors. Capturing such correlations is possible in the hierarchical Bayesian framework by introducing parametric relationships between the hyper-parameters and environmental factors. This issue is well addressed and demonstrated in [48], where the hyper-parameters are first updated and then trained on the environmental parameters, excitation amplitude, etc. This approach is also applicable to the method presented herein. However, this issue is out of the aims and scope of this study and will be studied in the future works. We also note that all experiments (laboratory and field experiments) were performed under very similar environmental conditions in an effort to demonstrate that the variability in the modal properties identified from different datasets is significant and larger than the identification precision. This variability is mainly due to modeling error, experimental variabilities, and data processing errors.

## 6. Concluding remarks and future work

Quantification of uncertainty in modal inference problems is revisited developing a new Hierarchical Bayesian modeling framework. This hierarchical model constitutes three major sub-models, namely the physics-based deterministic model, the prediction error model, as well as the hyper probability model. The variability of the dynamical parameters over multiple data sets/segments is captured through the hyper probability model, described using a multivariate Gaussian distribution. The mean and covariance of the hyper Gaussian distribution, referred to as the hyper-parameters, are unknown (uncertain) encapsulating the variability of the dynamical parameters over multiple data sets/segments. Subsequently, the Bayes' rule is adopted to compute the joint posterior distribution. The Laplace asymptotic approximation method is established to assess and simplify the likelihood function representing each data set. This approximation allows it to acquire analytical formulations for the hyper-parameters marginal distribution. The uncertainty characterized via the hyper-parameters is next propagated to compute a posterior predictive distribution for the dynamical parameters. This propagated uncertainty provides a holistic treatment of uncertainty capturing both the variability and identification precision. Since this task necessitates integrating over the hyper-parameters, a MCMC sampling algorithm (Algorithm 1) is proposed as a general computational approach. For a large number of data sets when the uncertainty of hyper-parameters can be neglected, the computation of the integrals can efficiently be carried out using another Laplace approximation. This approximation requires calculating the MAP estimation of the hyper-parameters, for which analytical derivatives and initial estimations are prescribed. This computing approach is outlined in Algorithm 2, called the dual asymptotic approximation approach. It is proved that only under some conditions the approximate MAP estimation of the hyper mean vector and the covariance matrix coincide with the ensemble mean vector and covariance matrix computed using the MPV corresponding to multiple data sets. Nevertheless, the proposed Bayesian framework is conceptually distinct, mathematically elegant, theoretically rigorous, which offers additional robustness with respect to different sources of uncertainty. This framework removes any need to incorporate inconsistent Frequentist concepts.

For a small number of data sets, it is often the case that the off-diagonal elements of the hyper covariance matrix are unidentifiable. This particularly prevails when the MAP estimations show sensitivity to the initial estimations, or when the MCMC sampler ends up with different chain of samples. To overcome this unidentifiability, the eignebasis decomposition of the hyper covariance matrix is established for separating the identification of eigenvalues from the eigenvectors. While the eigenvectors are set to reasonable estimations obtained from the approximate MAP estimations, the eigenvalues are suggested to be identified using the MCMC sampling, or the Laplace asymptotic approximation algorithms. Using this novel remedy not only preserves the identifiability of the parameters, it also considerably reduces the number of unknown parameters.

Ultimately, two experimental examples are employed to test and verify the proposed algorithms using the vibration measurements of a three-story prototype structure and a cable footbridge. It is demonstrated that the proposed framework effectively quantifies both the identification precision and the variability. This issue fills the existing gaps as to the quantification of the variability promoted due to modeling errors.

In this paper, the hierarchical Bayesian framework is presented for single setup data and well-separated resonant peaks. Efforts are underway to generalize it further for overlapping resonant peaks and multiple setup data, where the variability could be even more dominant as compared to the identification precision.

**Acknowledgements**

Financial support from the Hong Kong Research Grants Council under project No. 16234816, 16212918, and 16211019 is gratefully acknowledged. The third author also wishes to acknowledge financial support from the European Union's Horizon 2020 Research and Innovation Programme under the Marie Skłodowska-Curie grant agreement No 764547.

This paper is drawn from the PhD dissertation of the first author accomplished jointly at Hong Kong University of Science and Technology and Sharif University of Technology. The first author sincerely appreciates Professor Fayaz R. Rofooei for kind support and supervision at Sharif University of Technology.

**Appendix (A): Mathematical proof for Eq. (20)**

**Theorem-1**: Let $\mathbf{X} \in \mathbb{R}^n$ and $\mathbf{\mu} \in \mathbb{R}^n$ be two random vectors, and let $\mathbf{\mu}_0 \in \mathbb{R}^n$, $\mathbf{\Sigma}_0 \in \mathbb{R}^{n \times n}$, and $\mathbf{\Sigma} \in \mathbb{R}^{n \times n}$ denote known mean vector and covariance matrices. The following relationships hold for the multiplication of Gaussian distributions with common variables:

$$N(\mathbf{X} | \mathbf{\mu}_0, \mathbf{\Sigma}_0) N(\mathbf{X} | \mathbf{\mu}, \mathbf{\Sigma}) = N(\mathbf{\mu} | \mathbf{\mu}_0, \mathbf{\Sigma} + \mathbf{\Sigma}_0) N(\mathbf{X} | \mathbf{\zeta}, \mathbf{\Omega}) \tag{A1}$$

$$\int_{\mathbf{X}} N(\mathbf{X} | \mathbf{\mu}_0, \mathbf{\Sigma}_0) N(\mathbf{X} | \mathbf{\mu}, \mathbf{\Sigma}) d\mathbf{X} = N(\mathbf{\mu} | \mathbf{\mu}_0, \mathbf{\Sigma} + \mathbf{\Sigma}_0) \tag{A2}$$

where $\int_{\mathbf{X}} . d\mathbf{X}$ represents the integration $\int_{-\infty}^{+\infty} \int_{-\infty}^{+\infty} ... \int_{-\infty}^{+\infty} dx_1 ... dx_{n-1} dx_n$ for brevity, and the parameters are defined as

$$\mathbf{\Omega} = (\mathbf{\Sigma}_0^{-1} + \mathbf{\Sigma}^{-1})^{-1} \tag{A3}$$

$$\mathbf{\zeta} = (\mathbf{\Sigma}_0^{-1} + \mathbf{\Sigma}^{-1})^{-1} (\mathbf{\Sigma}_0^{-1} \mathbf{\mu}_0 + \mathbf{\Sigma}^{-1} \mathbf{\mu}) \tag{A4}$$

Proof of Theorem-1:

We first simplify the multiplication of Gaussian distributions in Eq. (A1). Due to the definition of multivariate Gaussian distributions, one can write:

$$N(\mathbf{X} | \mathbf{\mu}_0, \mathbf{\Sigma}_0) N(\mathbf{X} | \mathbf{\mu}, \mathbf{\Sigma}) = (2\pi)^{-n} |\mathbf{\Sigma}_0|^{-1/2} |\mathbf{\Sigma}|^{-1/2} \exp\left(-\frac{1}{2}\varphi\right) \tag{A5}$$

where $|.|$ denotes the matrix determinant, and $\varphi$ can be expressed as

$$\varphi = (\mathbf{X} - \mathbf{\mu}_0)^T \mathbf{\Sigma}_0^{-1} (\mathbf{X} - \mathbf{\mu}_0) + (\mathbf{X} - \mathbf{\mu})^T \mathbf{\Sigma}^{-1} (\mathbf{X} - \mathbf{\mu}) \tag{A6}$$

Since $\varphi$ is a scalar in this equation, the cyclic permutation property of the trace provides [69]:

$$\varphi = tr\left[ \mathbf{\Sigma}_0^{-1} (\mathbf{X} - \mathbf{\mu}_0)(\mathbf{X} - \mathbf{\mu}_0)^T + \mathbf{\Sigma}^{-1} (\mathbf{X} - \mathbf{\mu})(\mathbf{X} - \mathbf{\mu})^T \right] \tag{A7}$$

Here, $tr[.]$ denotes the trace of a matrix. After performing some basic algebraic operations, this expression can be rewritten as

$$\varphi = tr\left[ \mathbf{\Omega}^{-1} \mathbf{X} \mathbf{X}^T - 2 \mathbf{\Omega}^{-1} \mathbf{\zeta} \mathbf{X}^T + \mathbf{\lambda} \right] \tag{A8}$$

where the parameters $\mathbf{\Omega}$ and $\mathbf{\zeta}$ are introduced earlier in Eqs. (A3-A4), and $\mathbf{\lambda}$ is $\mathbf{\Sigma}_0^{-1} \mathbf{\mu}_0 \mathbf{\mu}_0^T + \mathbf{\Sigma}^{-1} \mathbf{\mu} \mathbf{\mu}^T$.

When we add and subtract the expression $\mathbf{\Omega}^{-1} \mathbf{\zeta} \mathbf{\zeta}^T$, the following quadratic form is obtained:

$$\varphi = tr\left[(\mathbf{X}-\boldsymbol{\zeta})^T \boldsymbol{\Omega}^{-1}(\mathbf{X}-\boldsymbol{\zeta})\right] + tr\left[\boldsymbol{\lambda} - \boldsymbol{\Omega}^{-1}\boldsymbol{\zeta}\boldsymbol{\zeta}^T\right] \tag{A9}$$

The next step is to simplify the expression $tr\left[\boldsymbol{\lambda} - \boldsymbol{\Omega}^{-1}\boldsymbol{\zeta}\boldsymbol{\zeta}^T\right]$. The cyclic permutation of the trace of three matrices gives:

$$tr\left[\boldsymbol{\lambda} - \boldsymbol{\Omega}^{-1}\boldsymbol{\zeta}\boldsymbol{\zeta}^T\right] = tr\left[\boldsymbol{\mu}_0\boldsymbol{\mu}_0^T\left(\boldsymbol{\Sigma}_0^{-1} - \boldsymbol{\Sigma}_0^{-1}(\boldsymbol{\Sigma}_0^{-1}+\boldsymbol{\Sigma}^{-1})^{-1}\boldsymbol{\Sigma}_0^{-1}\right) + \boldsymbol{\mu}\boldsymbol{\mu}^T\left(\boldsymbol{\Sigma}^{-1} - \boldsymbol{\Sigma}^{-1}(\boldsymbol{\Sigma}_0^{-1}+\boldsymbol{\Sigma}^{-1})^{-1}\boldsymbol{\Sigma}^{-1}\right)\right] \\ - 2tr\left[\boldsymbol{\mu}_0\boldsymbol{\mu}^T\boldsymbol{\Sigma}^{-1}(\boldsymbol{\Sigma}_0^{-1}+\boldsymbol{\Sigma}^{-1})^{-1}\boldsymbol{\Sigma}_0^{-1}\right] \tag{A10}$$

Due to the matrix inversion Lemma [69], the following equations hold:

$$(\boldsymbol{\Sigma}_0 + \boldsymbol{\Sigma})^{-1} = \boldsymbol{\Sigma}_0^{-1} - \boldsymbol{\Sigma}_0^{-1}(\boldsymbol{\Sigma}_0^{-1}+\boldsymbol{\Sigma}^{-1})^{-1}\boldsymbol{\Sigma}_0^{-1} = \boldsymbol{\Sigma}^{-1} - \boldsymbol{\Sigma}^{-1}(\boldsymbol{\Sigma}_0^{-1}+\boldsymbol{\Sigma}^{-1})^{-1}\boldsymbol{\Sigma}^{-1} \tag{A11}$$

$$(\boldsymbol{\Sigma}_0 + \boldsymbol{\Sigma})^{-1} = \boldsymbol{\Sigma}^{-1}(\boldsymbol{\Sigma}_0^{-1}+\boldsymbol{\Sigma}^{-1})^{-1}\boldsymbol{\Sigma}_0^{-1} \tag{A12}$$

Substituting these equations into the Eq. (A10) leads to the following quadratic from:

$$tr\left[\boldsymbol{\lambda} - \boldsymbol{\Omega}^{-1}\boldsymbol{\zeta}\boldsymbol{\zeta}^T\right] = tr\left[(\boldsymbol{\mu}-\boldsymbol{\mu}_0)^T(\boldsymbol{\Sigma}_0+\boldsymbol{\Sigma})^{-1}(\boldsymbol{\mu}-\boldsymbol{\mu}_0)\right] \tag{A13}$$

When this result is replaced into Eq. (A9), $\varphi$ is expressed as

$$\varphi = (\mathbf{X}-\boldsymbol{\zeta})^T\boldsymbol{\Omega}^{-1}(\mathbf{X}-\boldsymbol{\zeta}) + (\boldsymbol{\mu}-\boldsymbol{\mu}_0)^T(\boldsymbol{\Sigma}_0+\boldsymbol{\Sigma})^{-1}(\boldsymbol{\mu}-\boldsymbol{\mu}_0) \tag{A14}$$

Substituting $\varphi$ from this equation into Eq. (A5) yields:

$$N(\mathbf{X}|\boldsymbol{\mu}_0,\boldsymbol{\Sigma}_0)N(\mathbf{X}|\boldsymbol{\mu},\boldsymbol{\Sigma}) = (2\pi)^{-n}|\boldsymbol{\Sigma}_0|^{-1/2}|\boldsymbol{\Sigma}|^{-1/2} \times \exp\left(-\frac{1}{2}(\boldsymbol{\mu}-\boldsymbol{\mu}_0)^T(\boldsymbol{\Sigma}_0+\boldsymbol{\Sigma})^{-1}(\boldsymbol{\mu}-\boldsymbol{\mu}_0)\right) \\ \times \exp\left(-\frac{1}{2}(\mathbf{X}-\boldsymbol{\zeta})^T\boldsymbol{\Omega}^{-1}(\mathbf{X}-\boldsymbol{\zeta})\right) \tag{A15}$$

It is easy to show that the following relationship between the determinants holds:

$$|\boldsymbol{\Sigma}|^{-1/2}|\boldsymbol{\Sigma}_0|^{-1/2} = |\boldsymbol{\Sigma}_0+\boldsymbol{\Sigma}|^{-1/2}|\boldsymbol{\Omega}|^{-1/2} \tag{A16}$$

Substituting this equation into Eq. (A15) and using the definition of multivariate Gaussian distribution will prove Eq. (A1):

$$N(\mathbf{X}|\boldsymbol{\mu}_0,\boldsymbol{\Sigma}_0)N(\mathbf{X}|\boldsymbol{\mu},\boldsymbol{\Sigma}) = N(\boldsymbol{\mu}|\boldsymbol{\mu}_0,\boldsymbol{\Sigma}_0+\boldsymbol{\Sigma})N(\mathbf{X}|\boldsymbol{\zeta},\boldsymbol{\Omega}) \tag{A17}$$

In addition, Eq. (A2) can be proved as

$$\int_{\mathbf{X}} N(\mathbf{X}|\boldsymbol{\mu}_0,\boldsymbol{\Sigma}_0)N(\mathbf{X}|\boldsymbol{\mu},\boldsymbol{\Sigma})d\mathbf{X} = N(\boldsymbol{\mu}|\boldsymbol{\mu}_0,\boldsymbol{\Sigma}+\boldsymbol{\Sigma}_0)\int_{\mathbf{X}} N(\mathbf{X}|\boldsymbol{\zeta},\boldsymbol{\Omega})d\mathbf{X} = N(\boldsymbol{\mu}|\boldsymbol{\mu}_0,\boldsymbol{\Sigma}+\boldsymbol{\Sigma}_0) \tag{A18}$$

**Theorem-2**: An alternative form for the parameters $\boldsymbol{\Omega}$ and $\boldsymbol{\zeta}$ can be obtained using Kalman filter formulations giving:

$$\boldsymbol{\Omega} = \boldsymbol{\Sigma}_0 - \mathbf{K}_G \boldsymbol{\Sigma}_0 \tag{A19}$$

$$\boldsymbol{\zeta} = \boldsymbol{\mu}_0 + \mathbf{K}_G (\boldsymbol{\mu}_0 - \boldsymbol{\mu}) \tag{A20}$$

where $\mathbf{K}_G = \boldsymbol{\Sigma}_0 (\boldsymbol{\Sigma}_0 + \boldsymbol{\Sigma})^{-1}$ denotes the gain matrix. Note that the proof of this theorem can be found in [70,71].

**Corollary-1**: In Eq. (A2), substituting $\mathbf{X} \triangleq \boldsymbol{\lambda}_s$, $\boldsymbol{\mu}_0 \triangleq \hat{\boldsymbol{\lambda}}_s$, $\boldsymbol{\Sigma}_0 \triangleq \hat{\boldsymbol{\Sigma}}_{\lambda_s \lambda_s}$, and $\boldsymbol{\Sigma} \triangleq \boldsymbol{\Sigma}_{\lambda\lambda}$ gives:

$$\int_{\lambda_s} N\left(\boldsymbol{\lambda}_s | \hat{\boldsymbol{\lambda}}_s, \hat{\boldsymbol{\Sigma}}_{\lambda_s \lambda_s}\right) N\left(\boldsymbol{\lambda}_s | \boldsymbol{\mu}_\lambda, \boldsymbol{\Sigma}_{\lambda\lambda}\right) d\boldsymbol{\lambda}_s = N\left(\boldsymbol{\mu}_\lambda | \hat{\boldsymbol{\lambda}}_s, \boldsymbol{\Sigma}_{\lambda\lambda} + \hat{\boldsymbol{\Sigma}}_{\lambda_s \lambda_s}\right) \tag{A21}$$

This provides proof for the expression claimed earlier in Eq. (20).

**Corollary-2**: Considering the same notations as Corollary-1 and using Theorem-2 will provide:

$$N\left(\boldsymbol{\lambda}_r | \hat{\boldsymbol{\lambda}}_r, \hat{\boldsymbol{\Sigma}}_{\lambda_r \lambda_r}\right) N\left(\boldsymbol{\lambda}_r | \boldsymbol{\mu}_\lambda, \boldsymbol{\Sigma}_{\lambda\lambda}\right) \propto N\left(\boldsymbol{\lambda}_r | \hat{\boldsymbol{\lambda}}_r + \mathbf{K}_\lambda \left(\boldsymbol{\mu}_\lambda - \hat{\boldsymbol{\lambda}}_r\right), \hat{\boldsymbol{\Sigma}}_{\lambda_r \lambda_r} - \mathbf{K}_\lambda \hat{\boldsymbol{\Sigma}}_{\lambda_r \lambda_r}\right) \tag{A22}$$

where $\mathbf{K}_\lambda = \boldsymbol{\Sigma}_0 (\boldsymbol{\Sigma}_0 + \boldsymbol{\Sigma})^{-1}$ is the so-called gain matrix. This provides proof for Eq. (23).

**Appendix (B): Analytical derivation of the Cholesky decomposition**

Let $\mathbf{\Sigma} \in \mathbb{R}^{n \times n}$ be a positive-definite covariance matrix expressed as $\mathbf{\Sigma} = \mathbf{SRS}$, where $\mathbf{R} = [\rho_{ij}] \in \mathbb{R}^{n \times n}, \forall i,j = \{1,2,...,n\} : \rho_{ij} = \rho_{ji} \,\&\, \rho_{ii} = 1$ is the correlation matrix and $\mathbf{S} = \mathrm{diag}[\sigma_{ii}] \in \mathbb{R}^{n \times n}, i = \{1,2,...,n\}$ is a diagonal matrix containing the standard deviations $\sigma_{ii}$. The Cholesky factorization of the correlation matrix obtains $\mathbf{R} = \mathbf{LL}^T$, where $\mathbf{L} \in \mathbb{R}^{n \times n}$ is a unitary lower triangular matrix. Based on the formulation proved in [62], the matrix $\mathbf{L}$ can be expressed as

$$\mathbf{L} = \begin{bmatrix} 1 & 0 & 0 & \cdots & 0 \\ \rho_{12} & \sqrt{1-\rho_{12}^2} & 0 & \cdots & 0 \\ \rho_{13} & \dfrac{\rho_{23} - \rho_{12}\rho_{13}}{\sqrt{1-\rho_{12}^2}} & \sqrt{1 - \boldsymbol{\rho}_3 \mathbf{R}_2^{-1} \boldsymbol{\rho}_3^T} & \cdots & 0 \\ \vdots & \vdots & \vdots & \ddots & 0 \\ \rho_{1n} & \dfrac{\rho_{2n} - \rho_{12}\rho_{1n}}{\sqrt{1-\rho_{12}^2}} & \dfrac{\rho_{3n} - \boldsymbol{\rho}_3^{*n} \mathbf{R}_2^{-1} \boldsymbol{\rho}_3^T}{\sqrt{1 - \boldsymbol{\rho}_3 \mathbf{R}_2^{-1} \boldsymbol{\rho}_3^T}} & \cdots & \sqrt{1 - \boldsymbol{\rho}_n \mathbf{R}_{n-1}^{-1} \boldsymbol{\rho}_n^T} \end{bmatrix} \quad (B1)$$

where $\rho_{ij}$ denotes the (i,j) entry of the correlation matrix $\mathbf{R}$; $\boldsymbol{\rho}_i^{*j} \triangleq \begin{bmatrix} \rho_{1j} & \rho_{2j} & \cdots & \rho_{(i-1)j} \end{bmatrix}^T$ is a vector defined for $j \geq i$ and $\boldsymbol{\rho}_i \triangleq \boldsymbol{\rho}_i^{*i}$ is a special case of it; $\mathbf{R}_k = [\rho_{ij}]_{k \times k}$ is a partition matrix containing the first $k^{\text{th}}$ rows and columns of $\mathbf{R}$. Thus, the covariance matrix can be rewritten as $\mathbf{\Sigma} = \mathbf{SLL}^T\mathbf{S}$. This formulation preserves both the symmetry and the positive semi-definiteness conditions of the covariance matrix.

**Appendix (C): Computation of the analytical derivatives of $L(\boldsymbol{\mu}_\lambda, \mathbf{\Sigma}_{\lambda\lambda})$**

*C.1. Review from matrix algebra*

In this appendix, we derive the first and second derivatives of $L(\boldsymbol{\mu}_\lambda, \mathbf{\Sigma}_{\lambda\lambda})$ with respect to the hyper-parameters. To this end, we quickly review some important properties related to the derivatives of matrix and vector quantities from matrix algebra references [71,72], to which interested readers are referred. Let $\mathbf{X} \in \mathbb{R}^{N_x \times N_x}$ and $\mathbf{Y} \in \mathbb{R}^{N_x \times N_x}$ be two arbitrary invertible matrices, $\mathbf{a} \in \mathbb{R}^{N_x}$ be an invariable vector, and $\xi$ be a non-zero scalar. We also use $X_{mn}$ to denote the $(m,n)$ element of matrix $\mathbf{X}$. The determinant and trace are denoted by $\det(.)$ and $tr(.)$, respectively. Given these notation conventions, in general the following partial derivatives hold:

$$\frac{\partial \mathbf{X}}{\partial X_{mn}} \triangleq \mathbf{J}_{mn} = \begin{bmatrix} J_{ij} \end{bmatrix} \quad ; \quad J_{ij} = \begin{cases} 1 & ; \; i = m \text{ and } j = n \\ 0 & ; \; \text{Otherwise} \end{cases} \tag{C1}$$

$$\frac{\partial (\mathbf{Xa})}{\partial \mathbf{a}} = \mathbf{X} \tag{C2}$$

$$\frac{\partial (\mathbf{Xa})}{\partial \xi} = \frac{\partial \mathbf{X}}{\partial \xi} \mathbf{a} \tag{C3}$$

$$\partial \left( \ln |\det(\mathbf{X})| \right) = tr\left( \mathbf{X}^{-1} \partial \mathbf{X} \right) \tag{C4}$$

$$\partial tr(\mathbf{X}) = tr(\partial \mathbf{X}) \tag{C5}$$

$$\frac{\partial \left( \ln |\det(\mathbf{X})| \right)}{\partial \mathbf{X}} = \left( \mathbf{X}^{-1} \right)^T = \left( \mathbf{X}^T \right)^{-1} \tag{C6}$$

$$\frac{\partial (tr(\mathbf{a}^T \mathbf{X} \mathbf{a}))}{\partial \mathbf{a}} = tr\left( \mathbf{X}(\mathbf{a} + \mathbf{a}^T) \right) \tag{C7}$$

$$\frac{\partial (tr(\mathbf{a}^T \mathbf{X}^{-1} \mathbf{a}))}{\partial \mathbf{X}} = -\left( \mathbf{X}^{-1} \mathbf{a} \mathbf{a}^T \mathbf{X}^{-1} \right)^T \tag{C8}$$

$$\frac{\partial (\mathbf{X}^{-1})}{\partial \xi} = -\mathbf{X}^{-1} \frac{\partial (\mathbf{X})}{\partial \xi} \mathbf{X}^{-1} \tag{C9}$$

**Corollary-3:** The partial derivative of $\ln |\det(\mathbf{X} + \mathbf{Y})|$ with respect to $\mathbf{X}$ is calculated as

$$\frac{\partial \left( \ln |\det(\mathbf{X} + \mathbf{Y})| \right)}{\partial \mathbf{X}} = \left( (\mathbf{X} + \mathbf{Y})^{-1} \right)^T \tag{C10}$$

Proof

The derivative of $\ln |\det(\mathbf{X} + \mathbf{Y})|$ with respect to each element $X_{mn}$ is determined based on Eq. (C4) leading to

$$\frac{\partial \left( \ln |\det(\mathbf{X} + \mathbf{Y})| \right)}{\partial X_{mn}} = tr\left( (\mathbf{X} + \mathbf{Y})^{-1} \frac{\partial (\mathbf{X} + \mathbf{Y})}{\partial X_{mn}} \right) \tag{C11}$$

Due to Eq. (C1), the RHS can be rewritten as

$$\frac{\partial\left(\ln|\det(\mathbf{X}+\mathbf{Y})|\right)}{\partial X_{mn}} = tr\left((\mathbf{X}+\mathbf{Y})^{-1}\mathbf{J}_{mn}\right) \tag{C12}$$

Due to the definition of $\mathbf{J}_{mn}$ in Eq. (C1), when $(\mathbf{X}+\mathbf{Y})^{-1}$ is multiplied by $\mathbf{J}_{mn}$, it yields a matrix that its $n^{th}$ column is equal to the $m^{th}$ row of $(\mathbf{X}+\mathbf{Y})^{-1}$, while all other columns are zero. Consequently, the trace of this multiplication appearing in Eq. (C12) only selects the $(n,m)$ element of $(\mathbf{X}+\mathbf{Y})^{-1}$, and the derivative of $\ln|\det(\mathbf{X}+\mathbf{Y})|$ with respect to $X_{mn}$ is equal to $\left((X+Y)^{-1}\right)_{mn}$. This is equivalent to writing:

$$\frac{\partial\left(\ln|\det(\mathbf{X}+\mathbf{Y})|\right)}{\partial \mathbf{X}} = \left((\mathbf{X}+\mathbf{Y})^{-1}\right)^{T} \tag{C13}$$

**Corollary-4:** The partial derivative of $\mathbf{a}^{T}(\mathbf{X}+\mathbf{Y})^{-1}\mathbf{a}$ with respect to $\mathbf{X}$ is calculated as

$$\frac{\partial\left(\mathbf{a}^{T}(\mathbf{X}+\mathbf{Y})^{-1}\mathbf{a}\right)}{\partial \mathbf{X}} = -\left[(\mathbf{X}+\mathbf{Y})^{-1}\mathbf{a}\mathbf{a}^{T}(\mathbf{X}+\mathbf{Y})^{-1}\right]^{T} \tag{C14}$$

Proof

Similar to Corollary (3), we break this derivative into its elements. Since the expression $\mathbf{a}^{T}(\mathbf{X}+\mathbf{Y})^{-1}\mathbf{a}$ represents a scalar, we have:

$$\frac{\partial\left(\mathbf{a}^{T}(\mathbf{X}+\mathbf{Y})^{-1}\mathbf{a}\right)}{\partial X_{mn}} = \frac{\partial tr\left(\mathbf{a}^{T}(\mathbf{X}+\mathbf{Y})^{-1}\mathbf{a}\right)}{\partial X_{mn}} \tag{C15}$$

Based on Eq. (C5), the RHS of this equation can be simplified into

$$\frac{\partial\left(\mathbf{a}^{T}(\mathbf{X}+\mathbf{Y})^{-1}\mathbf{a}\right)}{\partial X_{mn}} = tr\left(\mathbf{a}^{T}\frac{\partial(\mathbf{X}+\mathbf{Y})^{-1}}{\partial X_{mn}}\mathbf{a}\right) \tag{C16}$$

where from Eq. (C9) it follows that:

$$\frac{\partial(\mathbf{X}+\mathbf{Y})^{-1}}{\partial X_{mn}} = -(\mathbf{X}+\mathbf{Y})^{-1}\frac{\partial(\mathbf{X}+\mathbf{Y})}{\partial X_{mn}}(\mathbf{X}+\mathbf{Y})^{-1} = -(\mathbf{X}+\mathbf{Y})^{-1}\mathbf{J}_{mn}(\mathbf{X}+\mathbf{Y})^{-1} \tag{C17}$$

Combining Eqs. (C16) and (C17) yields:

$$\frac{\partial \left(\mathbf{a}^T (\mathbf{X}+\mathbf{Y})^{-1}\mathbf{a}\right)}{\partial X_{mn}} = -tr\left(\mathbf{a}^T(\mathbf{X}+\mathbf{Y})^{-1}\mathbf{J}_{mn}(\mathbf{X}+\mathbf{Y})^{-1}\mathbf{a}\right) \qquad (C18)$$

The cyclic permutation of the trace of three matrices yields that $tr(ABC) = tr(CAB) = tr(BCA)$ [72]. Using this property provides:

$$\frac{\partial \left(\mathbf{a}^T (\mathbf{X}+\mathbf{Y})^{-1}\mathbf{a}\right)}{\partial X_{mn}} = -tr\left((\mathbf{X}+\mathbf{Y})^{-1}\mathbf{a}\mathbf{a}^T(\mathbf{X}+\mathbf{Y})^{-1}\mathbf{J}_{mn}\right) \qquad (C19)$$

Making the same argument as the one obtained earlier to derive Eq. (C13) ultimately gives:

$$\frac{\partial \left(\mathbf{a}^T (\mathbf{X}+\mathbf{Y})^{-1}\mathbf{a}\right)}{\partial X_{mn}} = -\left((X+Y)^{-1}\mathbf{a}\mathbf{a}^T(X+Y)^{-1}\right)_{mn} \qquad (C20)$$

This result suggests that the derivative of $\mathbf{a}^T(\mathbf{X}+\mathbf{Y})^{-1}\mathbf{a}$ with respect to $X_{mn}$ will be the corresponding element from the transpose of $-(\mathbf{X}+\mathbf{Y})^{-1}\mathbf{a}\mathbf{a}^T(\mathbf{X}+\mathbf{Y})^{-1}$. This therefore proves the formulation claimed earlier in Eq. (C14).

*C.2. Derivation of the gradient vector of $L(\mathbf{\mu}_\lambda, \mathbf{\Sigma}_{\lambda\lambda})$*

We first recall that the negative logarithm of the hyper-parameters' marginal distribution is given by

$$L(\mathbf{\mu}_\lambda, \mathbf{\Sigma}_{\lambda\lambda}) = \frac{1}{2}\sum_{s=1}^{N_D}\ln\left|\det\left(\mathbf{\Sigma}_{\lambda\lambda} + \hat{\mathbf{\Sigma}}_{\lambda_s\lambda_s}\right)\right| + \frac{1}{2}\sum_{s=1}^{N_D}\left[(\mathbf{\mu}_\lambda - \hat{\mathbf{\lambda}}_s)^T(\mathbf{\Sigma}_{\lambda\lambda} + \hat{\mathbf{\Sigma}}_{\lambda_s\lambda_s})^{-1}(\mathbf{\mu}_\lambda - \hat{\mathbf{\lambda}}_s)\right] + c' \qquad (C21)$$

The gradient of this function is composed of the derivatives of $L(\mathbf{\mu}_\lambda, \mathbf{\Sigma}_{\lambda\lambda})$ with respect to the elements of the hyper mean vector and covariance matrix. Due to (C7), the first derivative of this function with respect to the elements of $\mathbf{\mu}_\lambda$ is obtained as

$$\frac{\partial L(\mathbf{\mu}_\lambda, \mathbf{\Sigma}_{\lambda\lambda})}{\partial \mathbf{\mu}_\lambda} = \sum_{s=1}^{N_D}\left[(\mathbf{\Sigma}_{\lambda\lambda} + \hat{\mathbf{\Sigma}}_{\lambda_s\lambda_s})^{-1}(\mathbf{\mu}_\lambda - \hat{\mathbf{\lambda}}_s)\right] \qquad (C22)$$

In addition, the derivatives of $L(\mathbf{\mu}_\lambda, \mathbf{\Sigma}_{\lambda\lambda})$ with respect to the elements of $\mathbf{\Sigma}_{\lambda\lambda}$ can be computed from

$$\frac{\partial L(\mathbf{\mu}_\lambda, \mathbf{\Sigma}_{\lambda\lambda})}{\partial \mathbf{\Sigma}_{\lambda\lambda}} = \frac{1}{2}\sum_{s=1}^{N_D}\frac{\partial\left(\ln\left|\det\left(\mathbf{\Sigma}_{\lambda\lambda} + \hat{\mathbf{\Sigma}}_{\lambda_s\lambda_s}\right)\right|\right)}{\partial \mathbf{\Sigma}_{\lambda\lambda}} + \frac{1}{2}\sum_{s=1}^{N_D}\frac{\partial\left[(\mathbf{\mu}_\lambda - \hat{\mathbf{\lambda}}_s)^T(\mathbf{\Sigma}_{\lambda\lambda} + \hat{\mathbf{\Sigma}}_{\lambda_s\lambda_s})^{-1}(\mathbf{\mu}_\lambda - \hat{\mathbf{\lambda}}_s)\right]}{\partial \mathbf{\Sigma}_{\lambda\lambda}} \qquad (C23)$$

Based on Corollaries 1 and 2, the following equations holds:

$$\frac{\partial\left(\ln\left|\det\left(\boldsymbol{\Sigma}_{\lambda\lambda}+\hat{\boldsymbol{\Sigma}}_{\lambda_s\lambda_s}\right)\right|\right)}{\partial \boldsymbol{\Sigma}_{\lambda\lambda}} = (\boldsymbol{\Sigma}_{\lambda\lambda}+\hat{\boldsymbol{\Sigma}}_{\lambda_s\lambda_s})^{-1} \qquad (C24)$$

$$\frac{\partial\left[(\boldsymbol{\mu}_\lambda-\hat{\boldsymbol{\lambda}}_s)^T(\boldsymbol{\Sigma}_{\lambda\lambda}+\hat{\boldsymbol{\Sigma}}_{\lambda_s\lambda_s})^{-1}(\boldsymbol{\mu}_\lambda-\hat{\boldsymbol{\lambda}}_s)\right]}{\partial \boldsymbol{\Sigma}_{\lambda\lambda}} = -(\boldsymbol{\Sigma}_{\lambda\lambda}+\hat{\boldsymbol{\Sigma}}_{\lambda_s\lambda_s})^{-1}(\boldsymbol{\mu}_\lambda-\hat{\boldsymbol{\lambda}}_s)(\boldsymbol{\mu}_\lambda-\hat{\boldsymbol{\lambda}}_s)^T(\boldsymbol{\Sigma}_{\lambda\lambda}+\hat{\boldsymbol{\Sigma}}_{\lambda_s\lambda_s})^{-1}$$

(C25)

Note that in these formulations the symmetry of the covariance matrices is postulated. Substituting these analytical derivatives into Eq. (C23) yields:

$$\frac{\partial L(\boldsymbol{\mu}_\lambda,\boldsymbol{\Sigma}_{\lambda\lambda})}{\partial \boldsymbol{\Sigma}_{\lambda\lambda}} = \frac{1}{2}\sum_{s=1}^{N_D}\left[\left(\boldsymbol{\Sigma}_{\lambda\lambda}+\hat{\boldsymbol{\Sigma}}_{\lambda_s\lambda_s}\right)^{-1} - \left(\boldsymbol{\Sigma}_{\lambda\lambda}+\hat{\boldsymbol{\Sigma}}_{\lambda_s\lambda_s}\right)^{-1}\left(\boldsymbol{\mu}_\lambda-\hat{\boldsymbol{\lambda}}_s\right)\left(\boldsymbol{\mu}_\lambda-\hat{\boldsymbol{\lambda}}_s\right)^T\left(\boldsymbol{\Sigma}_{\lambda\lambda}+\hat{\boldsymbol{\Sigma}}_{\lambda_s\lambda_s}\right)^{-1}\right] \qquad (C26)$$

Based on Eqs. (C22) and (C26), it is straightforward to compute the gradient vector of $L(\boldsymbol{\mu}_\lambda,\boldsymbol{\Sigma}_{\lambda\lambda})$.

*C.3. Derivation of the Hessian matrix of $L(\boldsymbol{\mu}_\lambda,\boldsymbol{\Sigma}_{\lambda\lambda})$*

The second derivative of $L(\boldsymbol{\mu}_\lambda,\boldsymbol{\Sigma}_{\lambda\lambda})$ with respect to the elements of $\boldsymbol{\mu}_\lambda$ is followed from Eqs. (C2) and (C22) giving:

$$\frac{\partial^2 L(\boldsymbol{\mu}_\lambda,\boldsymbol{\Sigma}_{\lambda\lambda})}{\partial \boldsymbol{\mu}_\lambda^T \partial \boldsymbol{\mu}_\lambda} = \sum_{s=1}^{N_D}(\boldsymbol{\Sigma}_{\lambda\lambda}+\hat{\boldsymbol{\Sigma}}_{\lambda_s\lambda_s})^{-1} \qquad (C27)$$

To compute other elements of the Hessian matrix, it is desirable to drive the second derivatives with respect to the elements. Let $\Sigma_{\lambda\lambda,mn}$ denote the (*m,n*) element of the hyper covariance matrix $\boldsymbol{\Sigma}_{\lambda\lambda}$. The derivatives of $\partial L(\boldsymbol{\mu}_\lambda,\boldsymbol{\Sigma}_{\lambda\lambda})/\partial\boldsymbol{\mu}_\lambda$ with respect to $\Sigma_{\lambda\lambda,mn}$ can be computed as

$$\frac{\partial^2 L(\boldsymbol{\mu}_\lambda,\boldsymbol{\Sigma}_{\lambda\lambda})}{\partial \Sigma_{\lambda\lambda,mn}\partial \boldsymbol{\mu}_\lambda} = \sum_{s=1}^{N_D}\frac{\partial(\boldsymbol{\Sigma}_{\lambda\lambda}+\hat{\boldsymbol{\Sigma}}_{\lambda_s\lambda_s})^{-1}}{\partial \Sigma_{\lambda\lambda,mn}}(\boldsymbol{\mu}_\lambda-\hat{\boldsymbol{\lambda}}_s) \qquad (C28)$$

and the derivatives of $\partial L(\boldsymbol{\mu}_\lambda,\boldsymbol{\Sigma}_{\lambda\lambda})/\partial\boldsymbol{\Sigma}_{\lambda\lambda}$ with respect to $\Sigma_{\lambda\lambda,mn}$ are determined as

$$\frac{\partial^2 L(\boldsymbol{\mu}_\lambda, \boldsymbol{\Sigma}_{\lambda\lambda})}{\partial \Sigma_{\lambda\lambda,mn} \partial \boldsymbol{\Sigma}_{\lambda\lambda}} = \frac{1}{2}\sum_{s=1}^{N_D}\left[\frac{\partial\left(\boldsymbol{\Sigma}_{\lambda\lambda}+\hat{\boldsymbol{\Sigma}}_{\lambda_s\lambda_s}\right)^{-1}}{\partial \Sigma_{\lambda\lambda,mn}} - \frac{\partial\left(\boldsymbol{\Sigma}_{\lambda\lambda}+\hat{\boldsymbol{\Sigma}}_{\lambda_s\lambda_s}\right)^{-1}}{\partial \Sigma_{\lambda\lambda,mn}}\left(\boldsymbol{\mu}_\lambda-\hat{\boldsymbol{\lambda}}_s\right)\left(\boldsymbol{\mu}_\lambda-\hat{\boldsymbol{\lambda}}_s\right)^T\left(\boldsymbol{\Sigma}_{\lambda\lambda}+\hat{\boldsymbol{\Sigma}}_{\lambda_s\lambda_s}\right)^{-1}\right]$$

$$-\frac{1}{2}\sum_{s=1}^{N_D}\left[\left(\boldsymbol{\Sigma}_{\lambda\lambda}+\hat{\boldsymbol{\Sigma}}_{\lambda_s\lambda_s}\right)^{-1}\left(\boldsymbol{\mu}_\lambda-\hat{\boldsymbol{\lambda}}_s\right)\left(\boldsymbol{\mu}_\lambda-\hat{\boldsymbol{\lambda}}_s\right)^T\frac{\partial\left(\boldsymbol{\Sigma}_{\lambda\lambda}+\hat{\boldsymbol{\Sigma}}_{\lambda_s\lambda_s}\right)^{-1}}{\partial \Sigma_{\lambda\lambda,mn}}\right] \quad (C29)$$

where the derivatives $\partial\left(\boldsymbol{\Sigma}_{\lambda\lambda}+\hat{\boldsymbol{\Sigma}}_{\lambda_s\lambda_s}\right)^{-1}/\partial \Sigma_{\lambda\lambda,mn}$ are computed based on Eqs. (C1) and (C9) leading to

$$\frac{\partial\left(\boldsymbol{\Sigma}_{\lambda\lambda}+\hat{\boldsymbol{\Sigma}}_{\lambda_s\lambda_s}\right)^{-1}}{\partial \Sigma_{\lambda\lambda,mn}} = -\left(\boldsymbol{\Sigma}_{\lambda\lambda}+\hat{\boldsymbol{\Sigma}}_{\lambda_s\lambda_s}\right)^{-1}\mathbf{J}_{mn}\left(\boldsymbol{\Sigma}_{\lambda\lambda}+\hat{\boldsymbol{\Sigma}}_{\lambda_s\lambda_s}\right)^{-1} \quad (C30)$$

Like the argument explained in derivation of Eq. (C13), this equation can be simplified into

$$\frac{\partial\left(\boldsymbol{\Sigma}_{\lambda\lambda}+\hat{\boldsymbol{\Sigma}}_{\lambda_s\lambda_s}\right)^{-1}}{\partial \Sigma_{\lambda\lambda,mn}} = -\left[\left(\boldsymbol{\Sigma}_{\lambda\lambda}+\hat{\boldsymbol{\Sigma}}_{\lambda_s\lambda_s}\right)^{-1}\right]_{mn}\left(\boldsymbol{\Sigma}_{\lambda\lambda}+\hat{\boldsymbol{\Sigma}}_{\lambda_s\lambda_s}\right)^{-1} \quad (C31)$$

Note that in derivation of this equation the symmetry of the covariance matrices is considered implicitly. Substituting this finding into Eqs. (C28) and (C29) gives:

$$\frac{\partial^2 L(\boldsymbol{\mu}_\lambda, \boldsymbol{\Sigma}_{\lambda\lambda})}{\partial \Sigma_{\lambda\lambda,mn} \partial \boldsymbol{\mu}_\lambda} = -\sum_{s=1}^{N_D}\left[\left(\boldsymbol{\Sigma}_{\lambda\lambda}+\hat{\boldsymbol{\Sigma}}_{\lambda_s\lambda_s}\right)^{-1}\right]_{mn}\left(\boldsymbol{\Sigma}_{\lambda\lambda}+\hat{\boldsymbol{\Sigma}}_{\lambda_s\lambda_s}\right)^{-1}(\boldsymbol{\mu}_\lambda-\hat{\boldsymbol{\lambda}}_s) \quad (C32)$$

$$\frac{\partial^2 L(\boldsymbol{\mu}_\lambda, \boldsymbol{\Sigma}_{\lambda\lambda})}{\partial \Sigma_{\lambda\lambda,mn} \partial \boldsymbol{\Sigma}_{\lambda\lambda}} = -\frac{1}{2}\sum_{s=1}^{N_D}\left[\left(\boldsymbol{\Sigma}_{\lambda\lambda}+\hat{\boldsymbol{\Sigma}}_{\lambda_s\lambda_s}\right)^{-1}\right]_{mn}\left(\boldsymbol{\Sigma}_{\lambda\lambda}+\hat{\boldsymbol{\Sigma}}_{\lambda_s\lambda_s}\right)^{-1}$$
$$+\sum_{s=1}^{N_D}\left[\left(\boldsymbol{\Sigma}_{\lambda\lambda}+\hat{\boldsymbol{\Sigma}}_{\lambda_s\lambda_s}\right)^{-1}\right]_{mn}\left(\boldsymbol{\Sigma}_{\lambda\lambda}+\hat{\boldsymbol{\Sigma}}_{\lambda_s\lambda_s}\right)^{-1}\left(\boldsymbol{\mu}_\lambda-\hat{\boldsymbol{\lambda}}_s\right)\left(\boldsymbol{\mu}_\lambda-\hat{\boldsymbol{\lambda}}_s\right)^T\left(\boldsymbol{\Sigma}_{\lambda\lambda}+\hat{\boldsymbol{\Sigma}}_{\lambda_s\lambda_s}\right)^{-1} \quad (C33)$$

When these partial derivatives are obtained, we will be able to rearrange them in a matrix format using the Kronecker product of matrices. By doing so, the derivatives are obtained as

$$\frac{\partial^2 L(\boldsymbol{\mu}_\lambda, \boldsymbol{\Sigma}_{\lambda\lambda})}{\partial \boldsymbol{\Sigma}_{\lambda\lambda} \partial \boldsymbol{\mu}_\lambda} = -\sum_{s=1}^{N_D}\left[\left(\boldsymbol{\Sigma}_{\lambda\lambda}+\hat{\boldsymbol{\Sigma}}_{\lambda_s\lambda_s}\right)^{-1}\otimes\left[\left(\boldsymbol{\Sigma}_{\lambda\lambda}+\hat{\boldsymbol{\Sigma}}_{\lambda_s\lambda_s}\right)^{-1}(\boldsymbol{\mu}_\lambda-\hat{\boldsymbol{\lambda}}_s)\right]\right] \quad (C34)$$

$$\frac{\partial^2 L(\boldsymbol{\mu}_\lambda, \boldsymbol{\Sigma}_{\lambda\lambda})}{(\partial \boldsymbol{\Sigma}_{\lambda\lambda})^2} = -\frac{1}{2} \sum_{s=1}^{N_D} \left[ \left( \boldsymbol{\Sigma}_{\lambda\lambda} + \hat{\boldsymbol{\Sigma}}_{\lambda_s \lambda_s} \right)^{-1} \otimes \left( \boldsymbol{\Sigma}_{\lambda\lambda} + \hat{\boldsymbol{\Sigma}}_{\lambda_s \lambda_s} \right)^{-1} \right]$$
$$+ \sum_{s=1}^{N_D} \left[ \left( \boldsymbol{\Sigma}_{\lambda\lambda} + \hat{\boldsymbol{\Sigma}}_{\lambda_s \lambda_s} \right)^{-1} \otimes \left[ \left( \boldsymbol{\Sigma}_{\lambda\lambda} + \hat{\boldsymbol{\Sigma}}_{\lambda_s \lambda_s} \right)^{-1} \left( \boldsymbol{\mu}_\lambda - \hat{\boldsymbol{\lambda}}_s \right) \left( \boldsymbol{\mu}_\lambda - \hat{\boldsymbol{\lambda}}_s \right)^T \left( \boldsymbol{\Sigma}_{\lambda\lambda} + \hat{\boldsymbol{\Sigma}}_{\lambda_s \lambda_s} \right)^{-1} \right] \right] \quad (C35)$$

where $\otimes$ denotes the Kronecker product of two matrices. Ultimately, the Hessian matrix can be characterized as the following block matrix:

$$\mathbf{H}(\boldsymbol{\mu}_\lambda, \boldsymbol{\Sigma}_{\lambda\lambda}) = \begin{bmatrix} \dfrac{\partial^2 L(\boldsymbol{\mu}_\lambda, \boldsymbol{\Sigma}_{\lambda\lambda})}{\partial \boldsymbol{\mu}_\lambda^T \partial \boldsymbol{\mu}_\lambda} & \dfrac{\partial^2 L(\boldsymbol{\mu}_\lambda, \boldsymbol{\Sigma}_{\lambda\lambda})}{\partial \boldsymbol{\Sigma}_{\lambda\lambda} \partial \boldsymbol{\mu}_\lambda} \\ \dfrac{\partial^2 L(\boldsymbol{\mu}_\lambda, \boldsymbol{\Sigma}_{\lambda\lambda})}{\partial \boldsymbol{\mu}_\lambda^T \partial \boldsymbol{\Sigma}_{\lambda\lambda}} & \dfrac{\partial^2 L(\boldsymbol{\mu}_\lambda, \boldsymbol{\Sigma}_{\lambda\lambda})}{(\partial \boldsymbol{\Sigma}_{\lambda\lambda})^2} \end{bmatrix} \quad (C36)$$

where the constituting blocks are computed in Eqs. (C27), (C34), and (C35).